\documentstyle[12pt]{article}
\newcommand{\be}{\begin{equation}}
\newcommand{\ee}{\end{equation}}
\newcommand{\bea}{\begin{eqnarray}}
\newcommand{\eea}{\end{eqnarray}}
\newcommand{\beas}{\begin{eqnarray*}}
\newcommand{\eeas}{\end{eqnarray*}}
\newcommand{\ba}{\begin{array}}
\newcommand{\ea}{\end{array}}

\newcommand{\bfg}{\begin{figure}}
\newcommand{\efg}{\end{figure}}
\newcommand{\fr}{\frac}

\newcommand{\mb}{\mbox}
\newcommand{\lft}{\lefteqn}
\newtheorem{th}{Theorem}
\newtheorem{lm}{Lemma}
\newtheorem{cl}{Corollary}
\newtheorem{df}{Definition}
\newtheorem{E}{Example}
\newcommand{\bth}{\begin{th}}
\newcommand{\eth}{\end{th}}
\newcommand{\blm}{\begin{lm}}
\newcommand{\elm}{\end{lm}}
\newcommand{\bcl}{\begin{cl}}
\newcommand{\ecl}{\end{cl}}
\newcommand{\bdf}{\begin{df}}
\newcommand{\edf}{\end{df}}
\newcommand{\brk}{\begin{rm}}
\newcommand{\erk}{\end{rm}}
\newcommand{\hs}{\hspace}
\newcommand{\vs}{\vspace}

\newcommand{\vst}{\vspace*}
\newcommand{\lb}{\label}
\newcommand{\nl}{\newline}
\newcommand{\np}{\newpage}
\newcommand{\om}{\omega}

\newcommand{\al}{\alpha}

\newcommand{\ld}{\lambda}

\newcommand{\bib}{\bibitem}
\newcommand{\ct}{\cite}
\newcommand{\rf}{\ref}

\begin{document}
\title{\sc Quantum field theory as a problem of resummation\vs{0.4cm}\\
\protect\Large \it (Short guide to using summability methods)\vst{1cm} }
\author{\sc Alexander Moroz\thanks{Thesis submitted in partial fulfillment
of the
requirements for obtaining the degree of `Candidatus Scientarium'
(the Czechoslovak equivalent of PhD).}}
\date{\it Institute of Physics \v{C}SAV\\
\it  Na Slovance 2 \\
\it CS-180 40 Prague 8, Czechoslovakia}
\maketitle
\vst{3.9cm}
\begin{center}
{\large\sc Prague, July 1991}
\end{center}
\thispagestyle{empty}
\np
\pagenumbering{roman}
\tableofcontents
\np
\section{Introduction}
\lb{sec:int}
\subsection{Divergences of perturbation theory}
\lb{sub:div}
Very common situation which one encounters in physics
\pagenumbering{arabic}
is a lack of exact solutions. If $Z(g)$ is some physical quantity
considered as a function of a ``coupling
constant"${}$\footnote{Throughout the paper
we shall call
expansion parameter as a coupling. It may, however, correspond
to the inverse of physical coupling in the case of strong
coupling expansion, to $1/N$ in the case of $1/N$ expansion, etc.}
$g$ then the only  what we have frequently at disposal are
few first terms of perturbation theory,
$$Z(g) = a_o + a_1 g + a_2 g^2 + \ldots + a_n g^n .$$
Unfortunately, without any regularization this expansion is
badly defined  in quantum field theory.
In fact, in a formal perturbation expansion,
the coefficients $a_n$ are {\em divergent} and thus {\em undefined}.
This is first kind of divergences which, in  renormalizable
quantum field theories\footnote{including superrenormalizable ones},
can be cured by using a suitable
renormalization prescription which gives a
physically plausible recipe how to go
from undefined unrenormalized coefficients $a_n$ to the
renormalized ones. We shall denote the renormalized coefficients
by $a^R_n{}'s$. This means
that by renormalization one obtaines the perturbation series
in a renormalized coupling constant $g_R$,
$$Z^R (g_R) = a_o ^R + a_1 ^R g_R + a_2 ^R g^2_R
+ \ldots + a_n^R g^n_R,$$
with well defined coefficients $a_n^R{}'s$. In general, however, still one
kind of divergences remains -
it is the {\em divergence of the perturbation series as whole}.
Indeed,
if we do not consider the Yukawa model in two dimensions \cite{MS},
fermionic field theories with UV cutoff where, due to
the Pauli exclusion principle, the  radius of convergence of
perturbation theory for
normalized (connected) Green's function is
nonvanishing \cite{FMRS1}, as well as lattice gauge theories in the
strong coupling region \cite{OS},
perturbation theory of many other physical models, especially
without cutoffs and
 in the weak coupling region, is known to be {\em divergent}.
The latter divergence is the main subject of the thesis.
One has to know how to deal with such divergences, what is
their origin, their physical interpretation and significance
if one wishes to have logically consistent understanding of
quantum field theories.

In the case that a perturbation theory is {\em asymptotic} then
it is no problem to deal with it in the region of small
coupling (if we are only interested in perturbative phenomena).
This
is the case of asymptotically free theories at very large
momenta or weak coupled theories at low momenta, respectively.
If we are
interested in nonperturbative phenomena then  we stuck with
the problem to give a meaning to divergent series.
If the series is asymtotic (see section \rf{sub:as}) then
by optimalizing the bound on the rest term (\rf{oi1}) one gets the optimal
number of terms to be kept. By keeping either more or  less
terms results in that our prediction based on the perturbation
theory is {\em getting
worse}. This optimalization of the bound on the rest term
is for example essential to prove majority of our results.
Indeed, our proofs follow mostly the same scheme  and are in this
sense a bit boring (sometimes were also boring for
the author himself) : using the Euler-Maclaurin sum formula
for a given sum, its asymptotic evaluation by the saddle
point technique, and optimalization of the rest term.
Nevertheless the scheme turns out to be quite efficient.

In field theory the perturbation
series  cannot be truncated painless.
When  truncated perturbation theory becomes to depend on
unphysical (renormalization scheme dependent) parameters. This freedom
may be  again used to optimalize the perturbative prediction
and to extract a scheme
independent result \cite{1,2}.
Thus if a perturbation series is known to be asymptotic and divergent
some time one can get  a reliable prediction by keeping only
finite number of its terms. However to establish the number of terms
to be kept one needs to know the large order behaviour of
perturbation theory.
Information about large orders
ceases to be only perturbative information
and, in fact, it is a {\em nonperturbative} one.
The  fact that a presence of nonperturbative solutions can be seen by the
large order behaviour of  the  perturbation theory
could seem to be a bit surprising at a first moment
but perturbation theory, if (whenever it is possible) suitably
modified, is known to give even a
good,  nonperturbative approximation \cite{48}.
As recently shown
in \ct{Za} large orders of perturbation theory also determine
{\em multiparticle cross sections at asymptotically high energies}.
In order that this
          information be useful one needs a suitable
{\em summability method}. This is the aim of this thesis to provide
a selfconsistent treatment of new result of the author on
regular analytic summability method which have been published
in his several articles \ct{Mo,M,M1,M2}.

If one goes back to the history one finds that investigation of
convergence of perturbation
series in physical models was initiated by Dyson \cite{3}, who
considered the case of QED. More rigorous studies of the
problem have been then done by a number of authors. The divergence
of the pertubation series of the scalar bosonic
$\lambda\phi _d^3$ model was proved by \cite{4}.
More general polynomial bosonic  field
models  and  their  correct large order behaviour have been
studied, e.g., by \cite{5,BGZ}. Perturbation theory for the Yukawa
interaction (without cutoff's)
in $\ell$ dimensions  ($2<\ell<4$) among fermions was shown
to be divergent in  \cite{P}.  So far the divergence of the
perturbation theory has  been also proved for many quantities
in  other  models  of  quantum mechanics and quantum  field
theory including the energy-levels of an arbitrary
anharmonic oscillator \cite{9,GGS}, the ground state energy of
some nonpolynomial
oscillator \cite{AM1}, one-loop effective lagrangian  of
QED \cite{11}, Green's functions in QCD \cite{'tH1} and
recently also some amplitudes of the bosonic string theory \cite{GP}.

As for the explicite large order behaviour majority of results
are based on the Lipatov argument \ct{5}, which gives in general
the large order behaviour (for bosonic field theories) of the
following type,
\begin{equation}
\mid a_n \mid\leq A a^n n^\alpha (1+o(n))^n n!\ .
\lb{m6}
\end{equation}
The bound  was  obtained  by  a  formal
steepest descent method in the Euclidean path integral
formulation of quantum field theory for a variety of  models
\cite{5,BGZ,P,BGZ1,CO}.
Unfortunately, the method is  based  on some additional assumptions
which have not been proved
\cite{55} (cf. \cite{56})\footnote{For  quantum  mechanical models
as well as for models where  horn-shaped singularities are  absent
works  rather well.})
The result, obtained by a phase-space  expansion confirms
the Lipatov analysis in $<4$  dimensions, but in $4$ dimensions
renormalization was  shown to disturb the Lipatov behaviour
in such a way that the constant  $a=(3/2\pi ^2)$ should be
replaced by
$-\beta_1 /2$, where $\beta_1 (N)=(N+8)/(2\pi^2)$ is  the
first nonzero coefficient of the $\beta $-function. Note that the
ratio  $a/\beta_1$  tends to 0 as $N\rightarrow\infty$ \cite{60,61}.
Rigorous large order behaviour of the massless
$\phi_4^4$ theory  has  been  studied in
\cite{FMRS}, but only for an UV-cutoffed theory.
The   rigorous   large   order   behaviour   has  been only
established  in the case of a  $N$  component massive
$\lambda\phi _4^4$ model. It is worthwhile to mention that the large
order behaviour of bosonic string theory also differs from the
Lipatov behaviour. Here the $n$-th order of perturbation theory is of
magnitude $(2n)!$ in place of $n!$ in (bosonic) field
theories. For further details see \ct{Gr,MO,GM,GZ} and references
therein. It is worth remembering that
the factorial  growth of perturbation theory was proved
especially for bosonic theories while inclusion of fermions
slow the divergence  down (see, e.g., \cite{FMRS1,P}).
Moreover,  sometimes the factorial growth is an artefact of
the approximation used and a more  careful  analysis  may  give
even a convergent result \cite{C}.
Similar divergences one also encounters in general relativity
in connection with an existence of the de Witt integral \ct{Br}.

Due  to  the  field theoretical approach to
critical phenomena \cite{14}  initiated by Wilson \cite{15}
analogous divergences inevitably appear  in  the  study of
variety of statistical systems.
As for quantum spin systems with finite interaction
the situation with regard to convergence is a bit better, at least
in the high-temperature
region which corresponds to the strong coupling region of lattice
field theories \ct{OS,NR}.
This is due to the fact that lattice models
with finite interaction can be
transformed on a {\em polymer system} \ct{GJS}. Afterwords, by
using general results on polymer expansions \ct{GKu,KP}, one can
derive {\em convergent cluster expansions} for them \ct{OS,GJS,Sei}.
The convergence of cluster expansions allows to deduce upper and
lower bounds on expectations of various types of observables
such as Wilson loops, 't Hooft loops and others from bounds
on polymer activities. The convergence also implies exponential
decay of correlations as well as that above some temperature
(coupling) there is no phase transition.
This
expansion is not, however, a power series expansion in general. In
contrast to continuum field theory where one has general
arguments for its large order behaviour \ct{5} there is no
general
argument how single terms of the cluster expansion should behave.
In regard to convergent expansions in field theory and statistical
systems see also \ct{MP,Po,T} (and references therein).
There a convergent perturbation expansion
is  obtained if, instead of the  expansion in
renormalized  coupling  constant, the expansion  in
powers of the {\em running coupling constant} is used.

For  complexity we note that some quantities of topological
origin in superrenormalizable
theories, like topological mass term in planar $QED$ with the
Chern-Simons term, can have {\em finite} perturbation series expansion
with all  coefficients but the first two being zero.
This is ensured by the
so-called {\em non-renormalization theorem} \cite{CH}.

\subsection{Other perturbative techniquess}
\lb{sec:pt}
In
the last two decades a hard work has  been  done in
developing perturbation theory where perturbation
parameter is rank of the
symmetry group of an underlying  model.  This  approach was
originally developed by 't Hooft \cite{64}. He also noted a
connection between the $1/N$ expansion
and dual models \cite{65}.
If a connection were solidly  established, many of the leading
mysteries of QCD would  be solved. The dual model has built
into it confinement, with  quarks  at  the end of a string.
Also, a  clear  connection  between  QCD and the dual model
would mean  that  the  problem of dynamical mass generation
had been solved,  since the dual model certainly has a mass
scale (the Regge slope).

To elucidate the main features of the approach let us
consider the familiar Hamiltonian of the hydrogen atom:
          $$H = \frac{p^2}{2m} - \frac{e^2}{r} \cdot $$
          Since  for
          $e^2 = 0$ we can solve this problem exactly - it is
          simply the problem of the motion of a free particle - one's
          first hope  might be to try to understand the hydrogen atom
          in  the  small $e^2$ regime by treating the interaction  term,
$-e^2 /r$, as a perturbation. This  hope  is frustrated because in the
hydrogen atom
          $e^2$ is not really a relevant parameter. It can be
          eliminated
          from   the problem  by redefining the scale of distances.

          After a rescaling $r\rightarrow tr$, $p\rightarrow p/t$, with
          $t=1/me^2$, the Hamiltonian becomes
   $$H = (me^4)\left[\frac{p^2}{r} - \frac{1}{r}\right],$$
 and  one  sees that the ``coupling constant"
          $e^2$ appears only
          in the overall factor $me^4$ which serves merely to define
          the overall scale  of energies and which could be absorbed
          in a
          rescaling of the time coordinate. Therefore, except for
          the overall scale of  lengths and times, the physics of the
          hydrogen atom - and atomic and molecular physics in general -
          is independent of
          $e^2$, and perturbation theory in $e^2$ is meaningless.

          Atom and molecules can be described by the reduced
          Hamiltonian with
          $e^2$ scaled out. For the hydrogen atom the
          reduced Hamiltonian is
          $$H = \frac{p^2}{2} - \frac{1}{r} \cdot $$
          The reduced Hamiltonian contains no evident free parameter.
          However, without a free parameter there is no  perturbation
          expansion. Without a perturbation expansion what we can do?
Suppose, since this is the case of  QCD  and of many other
          analogous problems that we were unable  to  diagonalize the
          Hamiltonian exactly, and then even a computer solution were
          impractical. To make progress, we must make an expansion of
          some kind. Since there is no obvious expansion parameter we
          must find a hidden one. To find a hidden expansion
          parameter, we may treat as a free, variable parameter a
          quantity that one usually regards as given and fixed.

          For instance, we may take a cue from spectacular
          developments in the last years in
          critical phenomena. After decades in which the study of
          critical phenomena was
          frustrated by the absence of an expansion parameter,
          Wilson  and
          Fisher suggested that to introduce an  expansion parameter,
          one should regard the number of spatial dimensions not as a
          fixed number, three, but as a variable parameter
          \cite{66}. They
          showed that critical  phenomena  are  simple in four
          dimensions and that in
          $4-\varepsilon$  dimensions critical phenomena can be
          understood by perturbation theory in
          $\varepsilon$. Similar ideas were used by Bender at al.  to
          study  of $\lambda\phi _4^4$  theory and some
          aspects of stochastic quantization \cite{67}.

          How, by analogy, can we create an expansion parameter
          in our problem?
          Instead  of  studying  atomic  physics in three
          dimensions, where it possesses an
          $O(3)$ rotation symmetry we may
          consider atomic physics in $N$ dimensions,  so  that the
          symmetry  is  $O(N)$. One can show that atomic physics
          simplifies as $N\rightarrow\infty$ and can be
          solved for large
          $N$ by expansion in powers of $1/N$ \cite{68}.

          Although  this expansion in atomic physics is not very
          useful at  $N=3$ ($N$ must apparentely be at least six or
          seven for the
          $1/N$ expansion to give good result), illustrates the
          main  features of the approach. In QCD, as in  atomic
          physics, the coupling constant can be scaled out of the
          problem and the basic difficulty is
          the  same  as  in atomic
          physics  - {\em the seeming absence of an  expansion  parameter}.
          For further details see \cite{69}.

So far the $1/N$ expansion has been successfully used to
solve the Gross-Neveu model \cite{70}, in the study  of
two-dimensional Yang-Mills theory \cite{71}, $U(\infty)$ lattice
gauge theory \ct{Ka}, the generalized
two-dimensional $U(N)$ Thirring model \cite{72},
and the  $O(N)$-symmetric sigma model \cite{73}. In last years
results of $1/N$ expansion have been also succesfully applied to
two dimensional quantum gravity (see \ct{GZ} and references therein),
as well as to quantum antiferromagnets in connection with
high-temperature superconductivity \ct{Re}.
As for the study of planar field
theories ($N\rightarrow\infty$ limit) see \cite{24,74}.
The $1/N$ expansion can be improved by a suitable summability method
just as conventional perturbation theory \ct{FMR}.

Unfortunately, for Yang-Mills theory, the
$N=\infty$ limit does  not
seem to be exactly soluble, even though it is significantly
simpler.
Nonperturbative approches to Yang-Mills theory are
sorely needed in particle
physics. There  is  not  much  hope
that  the Yang-Mills theory itself is exactly  soluble,  so
one  has to look for another theory which is ``close enough"
to it.  Then  Yang-Mills theory can be studied by an
expansion around this soluble theory. As a general rule the
more symmetry a theory has the more easily it can be solved.

Can we enlarge the symmetry group futher and thus find
a soluble theory? The answer is ``yes".
Consider Yang-Mills theory with structure group $G$ on a
spacetime
$X\times R$. In the gauge $A_o =0$ the structure  group  is
the set $XG$ of all smooth maps from $X$ to $G$. Let us
consider $G=U(N)$. When we take the
$N\rightarrow\infty$ limit in  the  usual
way, we  consider  an  infinite sequence of Yang-Mills  theories
with structure groups

$$U(1)\subset U(2)\subset\ldots
U(N)\subset U(N+1)\subset\ldots U(\infty ).$$

The  gauge group of Yang-Mills theory in the
$N=\infty$ limit is then $XU(\infty )$, which contains
$XU(N)$ for any $N$. But there
exists a group (the universal gauge group)
$U_2$ which contains
$XU(N)$ not only for all $N$, but also for all $X$ \ct{75}.

There is too early to say whether this approach will be
successful, but it by no means provides a  useful  step  to
find an exactly soluble theory for perturbation theory
calculations.

\subsection{Asymptotic expansions and strong asymptotic conditions}
\lb{sub:as}
As it was discussed above perturbation series
in quantum theory are mostly divergent and can have at best
          the meaning of asymptotic series.
{\em Asymptoticity} means in general that the perturbation  theory cannot
determine the solution uniquely.
More precisely
given an arbitrary sequence $\{a_n\}_{n=o}^{\infty}$ of
complex numbers and an arbitrary sector-like domain $D$,
there  exists  for  some  $\varepsilon >0$ function
$f(z)$, which is regular in
\( D_{\varepsilon}:=  D\cap\{z\mid \mid z\mid <\varepsilon\}\)
and
\[\lim _{z\rightarrow 0, z\in D_\varepsilon} (f(z)-a_o-\ldots
-a_n z^n)/ z^{n+1} = a_{n+1}\]
exists, or equivalently
\begin{equation}
f(z)  \sim \sum _{n=o} ^{\infty } a_n z^n\hspace{1cm}
(z\rightarrow 0, z\in D_{\varepsilon }),
\label{h1}
\end{equation}
          i.e., $\sum _{n=o}^{\infty } a_n z^n $ is an asymptotic series
          of $f(z)$ in the region
          $D_{\varepsilon }$ \cite{R,H}. In general there are
          infinitely
 many functions with  the above properties.

Asymptotic series may be {\em divergent as well as convergent}
(note that a convergent asymptotic expansion does not prevent
a function from being singular at the origin).
To deal with such series one has to look for a maximal region $D$ of the
complex coupling constant
plane in which the asymptotic expansion is uniform. In order that such
series determine its sum $f(z)$  uniquely this series has to satisfy
some additional conditions, so called {\em strong asymptotic
conditions} (SAC). SAC are then
relations which put into the balance
the shape of the region $D$, or analyticity  in  the
  complex coupling constant plane, with some bound  on  the  rest term
          $R_N (z)$,
\be
f(z) = \sum _{n=o}^{N-1} a_n z^n + R_N (z),
\lb{oi1}
\ee
in such a way that only one function $f(z)$ can satisfy them.
SAC are conditions which by definition  {\em prevent} appearence of
nonperturbative terms like $\exp(-A/g)$ in perturbation theory
(or  a  typical ``tunnelling" like amplitudes),
i.e., terms whose perturbation expansion  is identically zero.
Indeed, to be unseen
in  the  standard perturbation theory the (non-zero)
contribution  of such a
term should have an asymptotic series which is identically zero.
Such are  for  instance  all nontrivial minima $A$ of
action $S$ for which
$S(A)<\infty$,  like {\em instantons} in  non-abelian  theories
\cite{BPST}  or  {\em kinks}  in  the  double-well anharmonic
potential \cite{Pl,BF}, etc.
Violation of SAC then indicates a
presence of nonperturbative effects in a theory
and {\em instability} of its ground state. It was emphasized that
the violation of SAC is a more serious problem  of a theory
than the divergence itself, therefore the {\em violation}  of SAC
leads to {\em ambiquity} of perturbation theory  predictions \ct{M2,S}.
Of course, this ambiguity depends on the behaviour of the rest term
$R_N(z)$ in (\rf{oi1}) for $N\rightarrow \infty$. As it has been exposed
above, optimalization of the bound may determine an optimal number
$N_o$ of terms to be kept and the value of $R_{N_o}(z)$ then determines
a maximal error by which the exact solution differs from the perturbative
one even in the case that the series violates SAC and does not
determine a unique solution.
Sometimes ambiguity of the perturbation
theory predictions, e.g., in bosonic string model are
highly  desirable, because ``if perturbative
string  theory  were make  sense,
string theory would have nothing to do with physical
reality" \cite{GP}. This is because there are many features
of perturbative treatment of string theory which are not shared by
the real world.
When  SAC are fulfilled then the coefficients $a_n$ of perturbation
theory
determine  $f(z)$ uniquely. To obtain this function an
appropriate analytic regular summability method can be used
\cite{H}. We say that a summability method ${\cal\sigma}$ is
{\em regular} if and only if
\begin{equation}
{\cal \sigma} (\sum _{n=o}^\infty a_n z^n )
=\sum _{n=o}^\infty a_n z^n ,
\label{h3}
\end{equation}
whenever the r.h.s. of (\ref{h3})
converges. Analogously, we say that  a  summability
method ${\cal \sigma}$ is {\em analytic} if for every power series
$\sum a_n z^n$ with nonzero radius of convergence
\begin{equation}
{\cal \sigma} (\sum _{n=o}^\infty a_n z^n ) = f(z) ,
\label{h4}
\end{equation}
whenever  the l.h.s. of (\ref{h4})
exists ($f(z)$ now being an analytic  continuation of
$\sum  a_n z^n$).
The  moment constant summability methods as defined by
\cite{H}  are automatically regular, although not  analytic
in general.
We confine
ourselves to the class of analytic moment constant summability method
(AMCSM) \cite{H},  one  member  of which is
the well-known  Borel summability method,
frequently used in physics (see \cite{M} for a recent review).

To illustrate the main features of the AMCSM let us consider a very
simple example of the series $\sum_{n=o}^\infty z^n$.
Using the identity
$$
\fr{1}{n!}\int_o^\infty e^{-t} t^n\ dt = 1,
$$
it is always true that
$$
\sum_{n=o}^\infty z^n = \sum_{n=o}^\infty \fr{z^n}{n!}
\int_o^\infty e^{-t} t^n \ dt,
$$
within the radius of convergence of $\sum_{n=o}^\infty z^n$, i.e.,
within the disc $\| z\|<1$.
However within the disc of convergence one can always interchange
summation and integration. Thus,
\be
\sum_{n=o}^\infty z^n = \int_o^\infty e^{-t}(\sum_{n=o}^\infty
\fr{(zt)^n}{n!})\ dt.
\lb{pic}
\ee
Here, the r.h.s. of (\rf{pic}) is nothing but the Borel sum of
$\sum_{n=o}^\infty z^n$. In this case it can be calculated explicitely,
\be
\int_o^\infty e^{-t} \fr{(zt)^n}{n!} \ dt =\int_o^\infty e^{-t}
e^{zt}\ dt = \int_o^\infty e^{-t(1-z)}\ dt =\fr{1}{1-z},
\ee
whenever $Re\, z<1$. Thus, we have obtained the expression which
converges for $Re\, z<1$, i.e., outside the disc of convergence of
$\sum_{n=o}^\infty z^n$
and gives there an analytic continuation of $\sum_{n=o}^\infty z^n$
(see Fig. \rf{fg3}).

\subsection{The Borel summability method}
\lb{sub:bor}
\subsubsection{The Nevanlinna theorem}
\lb{sub2:nev}
A main mathematical tool for investigation whether a series is Borel
summable or not provides the Nevanlinna theorem
\cite{N,So}:\vst{0.3cm}\nl

\bth : Let $f(z)$ be analytic in the circle
          \(C_R := \{ z\mid Re 1/z > 1/R \}\), continuous up to the
          boundary, and satisfy there the estimates
           \[f(z) =  \sum_{k=o}^{N-1} a_k z^k + R_N (z) ,\]
          with
          \begin{equation}
          |R_N (z)| \leq  A\sigma^N N!\mid z\mid^N  ,
\label{h7}
          \end{equation}
          uniformly in $N$ and in $z\in\bar{C_R}$. Then $B(t)$,
\[ B(t) = \sum _{n=o}^\infty a_n \fr{t^n}{n!},\]
          converges for $\mid t\mid<1/\sigma$ and has an analytic
          continuation  to  the  striplike region
          \(S_{\sigma}:=\{t\mid
          dist(t,R_+ ) < 1/\sigma\},\)
          satisfying the bound
          \begin{equation}
          \mid B(t)\mid\leq K e^{\mid t\mid /R}
          \label{h8}
\end{equation}
          uniformly  in  every  \(S_{\sigma '}\) with
          $\sigma '>\sigma$.  Furthermore, f can be represented by
          the absolutely convergent integral
\begin{equation}
          f(z) = (1/z)\int _o^\infty e^{-t/z} B(t) dt
\label{h9}
\end{equation}
          for any $z\in C_R$.

          Conversely,  if  $B(t)$  is a function analytic in
          $S_{\sigma ''}$ $(\sigma''<\sigma)$ and satisfying there
          the bound (\ref{h8}), the function $f(z)$
          defined by (\ref{h9}) is analytic in $C_R$ and satisfies
 (\ref{h7}) [with
          $a_n = B^{(n)}(t)\mid _{t=o}$] uniformly  in  every  $C_{R'}$
          with $R'<R$.
\lb{hth1}
\eth

\subsubsection{Application to physical models}
\lb{sub2:app}
          The reason why the Borel transform has been so successfully
          used in physics is due to the fact that the  path  integral
          can be rewritten in the Borel form \cite{'tH1,18}. Indeed,
          consider
          a euclidean functional integral which has been rescaled
          as in a loop expansion
     $$G(g^2) = \int [dA] \exp [-S(A)/g^2] .$$
          Functional  integrals  are  the $N\rightarrow\infty$ limit
          of integrals like
 $$ g^{-N}{\cal N}\int \prod _{i=o}^{N} \int
dx_i \exp [-S(x_i)/g^2] ,$$
          where  ${\cal N}$ is the usual normalizing factor. Using
          the Cauchy integral formula this can be rewritten as
$${\cal N}\int \prod _{i=o}^{N} dx_i \oint _C \frac{dz}{2\pi i}
          \exp (-z/g^2 )\frac{\Gamma (N/2+1)}{[S(x_i)-z]^{N/2+1}} ,$$
          where $C$ is  a contour encircling the range of $S$, which
          is normalized to run from 0 to $\infty$. An interchange of
          integrations
          is allowable  for  finite  $N$  and  reasonable actions
          yielding
          \begin{eqnarray}
          G(g^2) & = & \oint _C \frac{dz}{2\pi i}\exp(-z/g^2) B_N (z),
          \nonumber \\
 B_N (z) & = & {\cal N}\int\Pi _{i=o}^{N} dx_i \frac{\Gamma
         (N/2+1)}{[S(x_i)-z]^{N/2+1}}\,\cdot
\lb{m1}
          \end{eqnarray}
          A final step is to take functional integration limit,
          $B(z) = \lim_{N\rightarrow\infty} B_N (z)$. Note that since
          denominator is raised to power
          $N/2+1$,  naive  power  counting  guarantees convergence
          even when S is only quadratic (as in free theory).
          On the other hand the coefficients $a_n$ of the perturbation
          series  in  quantum  models do not  grow  faster   than
          $A\sigma ^n\Gamma (\alpha n  +\beta )$,  where $\Gamma$ is
          the usual gamma function and
          the  constants $A$, $\sigma$, $\alpha$  and $\beta$ do not
          depend on $n$, what is
          just  the  case which can be dealt with the Borel method.

          The  Borel  summability  of the perturbation series of
          the physical  quantities  has been studied by many authors.
          In quantum theory  it was firstly used by \cite{19} to show
          that
          one-loop  effective lagrangian of QED exactly calculated by
          Schwinger  in the proper-time formalism
\cite{20} is for the  vanishing
          electric field nothing but the Borel  sum  of  the
          perturbation series (see also
          \cite{21}). The Borel  summability
          of the perturbation series has been also proved in the case
of the energy-levels of an anharmonic oscillator in any
finite
dimension \cite{9,GGS,34}, the resonances of the Stark effect on a
Hydrogen-like atom \cite{21,22}, in the case of the Zeemann
effect
\cite{23}, for some nonpolynomial oscillators \cite{AM1}
and for
the  planar  asymptotically  free massive field theory  for
sufficiently small  coupling
\cite{24}. The method was also used
for a rigorous perturbative construction of the massive
euclidean
Gross-Neveu model \cite{FMRS1}. In combination with conformal
mapping or with the Pad\'{e} approximation the Borel
summability
has  been  successfully used to compute the critical
coefficients  of  the statistical models
\cite{26}, the zeros of
the beta  function \cite{27}, the ground state energy of the
anharmonic
oscillator  at  infinite  coupling  limit \cite{28}, to
compute the beta function of the  massive
$\lambda\phi _4^4$ model \cite{Ko} or
to sum dominant coupling constant logarithms in
$QED_3$ \cite{30}, etc.

\subsubsection{Structure of the Borel transform}
\lb{sub2:str}

In  physically  realistic  models the structure of the
Borel transform is very  involved. The obstacles for the
Borel summing of a perturbation series appear as
singularities
of the Borel transform on the real positive axis.
These properties of the Borel transform can be deduced as
follows.
The  integral  for  $B_N (z)$ is very much like a Feynman
parameter integral.  Thus  the Landau conditions may be
applied
to ask when $B_N (z)$ is singular \cite{SG}. The Landau
conditions
have a direct interpretation in terms of solutions to
the  euclidean  equations  of motion: $B_N (z)$ may be
singular  for $z$ such  that  there  exist  $x_i^s$ for which
 \begin{eqnarray}
          z &=& S(x_i^s) ,\nonumber\\
          0 &=& \partial S/\partial x_i \mid _{x_i=x_s} .
\lb{m2}
          \end{eqnarray}
This  means  $B_N (z)$  may be singular for $z$ equal to the
action of a  solution  to the classical euclidean equations
of motion (discretized).  This  is true for complex as well
as real-valued solutions. Such solutions are generally
called
{\em pseudoparticles} or {\em instantons}.  It  is  widely supposed
that in the large-$N$ limit the solutions  to the discretized
equations approach a limit which is identified with the
solution to the continuum equations of motion.

In  general,  attached  to  each singularity will be a
{\em branch cut}. Because  $B(z)$  is  integrated along a contour
$C$ encircling the positive real axis, it is convenient for real,
positive singular points to have their branch cuts
running along the positive real axis.

The  only  singularities on the first sheet are on the
positive  real axis. These can arise  from  real  solutions
with real positive action. When no instantons  are  present
the only singularity on the non-negative real $z$ axis is at
$z=0$, the singularity arising from the vacuum solution.
There is a cut extending from this singularity to positive
infinity. When instantons occur, there will be another
contributions to the Borel sum othe than those visible in
perturbations about
vacuum. The instanton singularities will
have branch  cuts attached to them. The total discontinuity
of $B(z)$ will be a sum of discontinuities across branch cuts.

In  renormalizable  (not  superrenormalizable)  theory
there are rather persuasing arguments that
apart from  instantons  another  singularities  can appear
which go under  generic  name  of  {\em renormalon} singularities
\cite{'tH1,35,Da}. A toy,  finite  dimensional model where
renormalon
singularities appear  has  been investigated in
\cite{36} in the
context  of  Wilson's  renormalization group acting in  the
space  of hamiltonians of the cutoff system. The renormalon
singularieties have been shown to be related to the
presence of a marginal direction at the fixed point of the
renormalization
group besides the relevant or  irrelevant  ones,
or to the {\em resonance} in the dynamical system language.
Despite great effort presence of renormalons in renormalizable
quantum field theories is not rigorously established yet \ct{60,61}.
This important unsolved problem is in a close connections
to the so-called {\em triviality problem} - all known four-dimensional
renormalizable quantum field theories are expected to be trivial
\ct{63,LW}.

The suggested structure of the Borel transform of QED
and  QCD  is shown in Fig. \rf{fg1a} and \rf{fg1b}.
\bfg
\vs{6cm}
\caption{{\em Supposed structure of the Borel transform of
QED. Here the units are $3\pi$ if $\alpha$ is the original expansion
parameter.}}
\lb{fg1a}
\efg
\bfg
\vs{6cm}
\caption{{\em Borel $z$ plane for QCD. The circles denote IR
divergences that might vanish or become unimportant in colour-free
channels.}}
\lb{fg1b}
\efg
Note that the position of
renormalon depends  on  the  sign  of the first nonvanishing
coefficient of  the  corresponding $\beta$-function. While
in an asymptotic free  theory UV-renormalons are harmless (lie on
the  negative  real axis) in nonasymptotic free theory like
QED renormalons  prevent  the Borel summability (lie on the
positive real axis). As for QED see also \cite{37}.
(We would  like  to  stress out that the calculations which
led to the renormalon \cite{'tH1,35} was only performed for
specific subclass of all graphs, which contribute to the given
order of a theory, and may only serve as indication but not
at all as a proof).

\subsubsection{Shortcomings of the Borel summability method}
\lb{sub2:short}

Unfortunately,  the  use of the Borel summability method is
sometimes very limited. Nonsummability means that
perturbation  theory  violates strong asymptotic
conditions (SAC), or, what is equivalent that perturbation theory
predictions {\em can not be unambigious}.
This happens due to presence of nonperturbative
effects which cause perturbation theory around the trivial minimum
to be unreliable.
Nonperturbative effects do appear and
destabilize a trial (classical) ground state in models having
classically degenerate
ground state \ct{BGZ2} which is the case, e.g., of the double well
potential \cite{Pl,BF,Z}, gauge
field theories \cite{BPST,Pl,BF}, as well as of the heterotic
string and superstring
\cite{DS}. However, SAC can be also violated
in the case of perturbation theory around a stable
nonperturbative ground state
(after the nonperturbative  effects
have removed the degeneracy of the  ground state \cite{Pl}),
as has been shown in the case of the  double-well potential
\cite{Z}\footnote{I thank J. Magnen for acquinting me with this
reference.}.
The physical meaning of the non Borel
summability is illustrated  on  some simple models in \cite{BGZ1,CO}.

The next shortcomings of the Borel summability method is connected
with its analytic properties. Indeed whenever the
Borel sum exists then it defines a function {\em regular} in some
{\em sector-like neighbourhood of the origin} (the convergence of
the Borel integral {\em need not} be absolute!) \cite{W}.
On the other hand a
maximal region of analyticity of realistic field theories in the
coupling
constant plane may be very small. For instance, there are strong
arguments that two-point Green's functions of four dimensional
renormalizable
massless field theories are only analytic in a {\em horn-shaped region}:
a wedge
bisected by the real axis and bounded  above and below by circles
which are tangent to the real  axis at the origin \cite{'tH1,K1,Wi}
(see Fig. \rf{fg2}).
\bfg
\vs{8cm}
\caption{{\em Expected  form of the maximal region of analyticity
in the coupling constant complex plane of four dimensional
renormalizable  field theories.}}
\lb{fg2}
\efg
Hence the maximal region of analyticity has a {\em cusp},
i.e., {\em zero opening angle at the origin}. The appearence of a
horn-shaped region of
analyticity is in fact a {\em nonperturbative result} which
we have about a theory. It is not seen in a perturbation theory
and arises if the {\em analytic structure} of the (response)
Green functions for complex momenta $p^2$
is combined with {\em asymptotic freedom}. This is the essence of
't Hooft's argument \cite{'tH1,Wi}
which relates the  standard  momentum space branch points for timelike $p^2>0$
to singularities in the complex plane of a suitably chosen coupling constant
by solving the Callan-Symanzik equation (for generalizations to more general
couplings see \cite{K0}). In fact, the use of the Callan-Symanzik
equation is not crucial in 't Hooft's argument and it goes through even if
a theoretical evidence of the asymptotic freedom is
replaced by an {\em experimental evidence} (if
available) of the asymptotic
freedom, i.e., that a measurable coupling  $g^2(\mu )$ behaves like
\begin{equation}
\frac{\mu d}{d\mu} g^2(\mu) = -\beta_1 g^4 (\mu) +{\cal O}(g^6 (\mu)),
\label{cc}
\end{equation}
on a sufficiently high mass scale $\mu$. Note that in this case we
always can write an analogue of the equation (\ref{cc}). Provided that
 such information is available one gets
rid of the problem of definition of a suitable coupling and the horn-shaped
region appears as UV effect once one assumes that at high momenta,
measurable quantities are only functions of the ratio $p^2/\mu_o ^2$,
$\mu_o$ being an integration constant of (\ref{cc}). Analytic
structure of response functions is by itself a
{\em nonperturbative information} about a theory which arises when very
general theoretical principles of unitarity and causality are combined
with experimental data. Hence, the appearance of horn-shaped regions is in
fact a very general phenomenon resulting
whenever the following points are satisfied or available: \vspace{0.2cm}\\
\noindent 1)  {\em unitarity}\\
\noindent 2) {\em causality}\\
\noindent 3) {\em asymptotic freedom} (either theoretical or
experimental evidence)\\
\noindent 4) {\em experimental data} (which determine the position of
thresholds and branch points in the momentum plane).\vspace{0.2cm}\\
Some other arguments for producing
singularities  in  the $g$-plane are  not  at all related to
branch points in the momentum  plane, and are due to the
{\em separatrix}
of the differential equation for the running coupling constant, i.e.,
to the line in the $g$-plane  which separates the asymptotically
free region  of
initial values $g$ from the non-asymptotically one \cite{KR,GK}. If the
Callan-Symanzik procedure for
calculating the asymptotic behaviour of one-particle-irreducible Green's
functions in the deep Euclidean region \cite{Sy} is correct then such
horn-shaped
regions of analyticity is a typical UV effect of all four dimensional
renormalizable field theories.
What is interesting is that a horn-shaped region is also the maximal
region
of analyticity  in the complex coupling constant plane of energy levels
of a simple quantum mechanical model - the anharmonic oscillator
in the massless limit \cite{Wi}.

If one imposes UV
cutoff then two- and four-point Green's functions of a four dimensional
massless field theory can be analytic in a
disc as was exemplified in a rigorous construction of the critical
$\phi ^4 _4$ theory \cite{FMRS}. However, if one removes this UV
cutoff
the disc should shrink to zero.  Recently it was shown that
some other quantities, like $\beta$-function
 (defined perturbatively using the BPHZ
renormalization with subtractions at vanishing external momentum)
of the massive $\phi ^4_4$ model without UV cutoff may be Borel summable
 and hence analytic in a disc \cite{Ko}. For the sake of completness we
recall that in the case of $\phi ^4_d $   with $d<4$ there is no  horn of
singularities \cite{Wi}.
As  for  the nongauge field theories some arguments in
favour  of  ambiguity of renormalized
$\phi _4^4$ field theory  has been given due to $1/N$ expansion \cite{59}.

The Lipatov bound (\rf{m6})
is compatible  with  SAC  only for a regions of the
asymptotic type $(0,\eta )$ (a is the same for all quantities
in a theory \cite{5,BGZ,P,BGZ1,CO}).
As we have mentioned above a general rigorous
justification of the Lipatov method has not been established so far.
However further investigations
would  not  change the fact that perturbation theory
of four dimensional renormalizable field theories violates
SAC. If  the large
order behaviour of the model in the limit of vanishing
UV-cutoff will be proved to violate (\rf{h10}) then this will
allow for nonperturbative solutions to the theory and
can sheds new  light  on
the triviality problem of the above theories \cite{63,LW}.
To get reliable  perturbation theory  one  must find a
nonperturbative ground state and
expand  around it \cite{57}. In the case of the double
well potential also some other techniquess were advocated
\cite{58}.
%Horn
However,  we cannot exclude the situation that
the  coefficients $a_n$ of perturbation theory will obey
the bound (\rf{h10}) in this model.
Provided  it is so than no nonperturbative
solutions can occur and the Borel transform should be an
entire function, what contradicts  expected singular
structure
of the  Borel  transform.  However, it may well happen that
there is  no  reason  to look at the Borel transform of the
four dimensional  renormalizable massless field theory,
because in this theory  convergence  of  the  Borel transform
contradicts the observed horn-shaped  region of analyticity
\cite{M,W}, in distinction from the lower  dimensional
case \cite{KR}.\np

\section{Goals and methods of the thesis}
\lb{sec:goal}
\subsection{Goals}
\lb{sub:goal}
The main goal of the thesis is to overcome some shortcomings of
the Borel summability method. Let us summarize the shortcomings
which have been solved by the author.
\vs{0.4cm}\\
 {\bf A}) {\em the Borel method cannot cope with the horn-shaped
singularity, exhibited in the massless QCD}
\cite{'tH1}, {\em in the
``massless"  limit of the anharmonic oscillator} \cite{Wi}, {\em and
in
some  other models. (For recent status of the  problem  see}
\cite{KR} {\em and references therein.)}\vs{0.4cm}\\

Another shortcoming of the Borel method is the following.
\vs{0.4cm}\\
{\bf B}) {\em In the regular case, i.e., when the series on the r.h.s.
of (\ref{h1}) has a nonzero radius of convergence, it can in
general
be analytically continued onto a region of the complex
plane which is larger than the region of $z$ for  which  the
Borel  sum exists. If the analytic continuation  $f(z)$  is
not an entire function, the Borel sum may not exist
for all $z$ from the Mittag-Leffler  (principal) star of
f(z)}
\cite{H} {\em (further $MLS(f)$),  because  always some
sector-like domain has to be discarded from the complex plane}
\cite{M,SG}.\vs{0.4cm}\\

To illustrate point ({\bf B}) note that the standard Borel
method sums the series $\sum _{n=o}^\infty z^n$ only in the
complex halfplane
$Re\, z<1$ but the analytic continuation $1/(1-z)$ of the
series  exists  in the whole complex plane except the point
$z=1$.
\bfg
\vs{8cm}
\caption{{\em Region of the Borel summability of $1/(1-z)$.}}
\lb{fg3}
\efg
We should like to demonstrate in simple examples that
the Borel summability method {\em does not have any priviliged role and
that it is just an ordinary method among a continuous
variety
of other analytic moment constant  summability methods (AMCSM)}. It
should be used when a divergent series has to
be summed up to a function regular in a disc $K(0,R)$. Provided the sum
of a
given divergent series is known to be analytic in some other
region then the corresponding  AMCSM
{\em compatible} with this analyticity requirement has to be used.

To generalize known SAC to
horn-shaped regions which may be of some physical importance is another
goal of this thesis.
The well-known example of SAC provides the Nevanlinna theorem
\cite{S,N}. These conditions are, however,
applicable only when sum of a given perturbation series is to be
analytic
in a disc $K(0,R):=\{z\mid Re(1/z)>1/R\}$  tangent to the imaginary
axis at the  origin (modulo mapping
$z\rightarrow z^{1/\alpha }$). On the other hand, we have seen that
many theories are known to be analytic in a horn-shaped region
only.
Without the validity of SAC one
{\em cannot} prove that a statement which is valid order by order in
the perturbation
theory is a property of the full theory. One such example is, e.g., the
Callan-Symanzik procedure where certain mass insertion terms are ignored
and assumed small even though one has been able to prove this fact order
by
order. It was found \cite{K1} that the Borel summability or SAC given in
\cite{H,S,N} cannot help in justifying the Callan-Symanzik procedure
just because of the horn-shaped singularities. In \cite{K1}
it was stated that
the horn-shaped singularity puts the Callan-Symanzik assumption about
the mass insertion term in doubt. Our point of view is that
there is {\em no controversy}
between the horn-shaped singularity and the Callan-Symanzik procedure
since one has to use SAC appropriate to a given analyticity region.

We shall also give a further improvement of the SAC as discussed in
\cite{S,N,So}, and show that
a product of functions obeying these SAC also obeys the
same SAC.
The knowledge about the relation of SAC and summability of
perturbation theory is important in order to
understand, e.g.,
the recent result on divergency of the bosonic string perturbation
theory  \cite{GP}  and frequently
quoted in connection with two dimensional quantum gravity
since there no care
is taken about analyticity to which the perturbation theory should be
summed up, and physical consequences are simply drawn from
Borel nonsummability.
Similar ``negligence" of analytic properties also occurs, e.g., in
\cite{BGZ,BGZ1} or \cite{MO}. Fortunately, the
series treated in \cite{BGZ,GP,BGZ1}  can be dealt
by Lemmas \rf{slm2} and \rf{slm3} (see below).
As for \cite{MO} the results of this paper can be valid
provided a fixed-angle elastic scattering amplitude of bosonic string
${\cal A}(s,\phi ;g)$ (see also \cite{GM}) is
analytic in a region $D$ of the sheeted complex plane
with an opening angle $\Theta\geq 9\pi$  since the resummed amplitude
${\cal A}_{\em resum}$ is such \cite{GGS,N}.

SAC discussed below may provide a tool
to decide whether such ambiguity  (degeneracy of the ground
state of an underlying theory) is  actually  present.
However, some futher work has to be accomplished  in  order to gain the
maximal  domain  of the analyticity, establish  a  rigorous
large order  behaviour or to give a suitable lower bound on
the coefficients of the perturbation theory.

\subsection{Methods}
\lb{sub:meth}

To solve the above shortcomings ({\bf A}) and ({\bf B}) of the
Borel method we use theory of complex variables \ct{SG}, asymptotic
series \ct{F,J}, and divergent series \ct{R,H}.
We  shall  start with the so-called regular case, i.e.,
when the  series on the r.h.s. of (\rf{h1}) has nonzero radius of
convergence. Then, by virtue of Theorem \rf{rth1}, our method will
sum $\sum _{n=o}^\infty z^n$  to $1/(1-z)$ in the whole
complex plane except the
ray $[1,\infty )$ (=$MLS[1/(1-z)]$)   in contrast to the
Borel method (see Fig. \rf{fg3}).

In the next we shall consider moment constant summability
method with the momemt sequence $\{\mu (n)\}_{n=o}^\infty$,
\be
\mu (n) := \int _o^\infty \exp (-\exp t)\ t^n\ dt,
\lb{m3}
\ee
i.e., with the moments in the {\em Stieltjes form}. Firstly
we shall prove that
the moment constant method solves the above difficulties ({\bf A})
and ({\bf B}) of the Borel method, and, in the next, we shall give
a whole family of such methods.
Indeed,  one of our main  results is that if $f(z)$ is
the principal  branch  of  an  analytic function regular at the
origin, where $f(z) =\sum a_n z^n$, then the integral
$I(z_o)$,
\[I(z_o) =  \int _o^\infty \exp (-\exp t) \sum _{n=o}^\infty
a_n \frac{(z_o t)^n}{\mu (n)}\, dt ,\]
converges if and only if $z_o\in MLS(f)$. If $z_o\in MLS(f)$ then
 \[f(z_o) = I(z_o).\]
Thus,  the  method is {\em analytic} and {\em regular}. The convergence
is  absolute and uniform in any bounded  subset  of $MLS(f)$
with nonzero distance from the boundary  of $MLS(f)$, and one
can also differentiate inside the sign of integration.
As for the property ({\bf B}) the method is analogous to the
Lindel\"{o}f one \cite{Mo}. In contrast to the Borel method
the method {\em does not only see beyond singularities} and may be
used for
localization of critical points of some physical
models. We shall also show that the issues  of  the  Borel
summability method consist in that it is based on the
asymptotic properties of  the  Mittag-Leffler  function
 approaching zero in the {\em complement of some sector with
nonzero opening angle} only \cite{H,E}, while the new method
(\rf{m3}) is based on an entire function which approaches zero
{\em in all radial directions except for one}.
In an application to the Rayleigh-Schr\"{o}dinger perturbation
theory, theory of linear operators in Banach and Hilbert spaces is used
\ct{K}.
\newpage
%REG
\section{New results}
\lb{sec:new}
\subsection{Regular case}
\lb{sub:reg}

{}From some general considerations we expect that if a
moment constant summability method exists
solving problems ({\bf A})  and
({\bf B}),  then  the moment sequence has to grow like
$(\ln n)^n$ (see
Remark  5). However, the problem of finding a nondecreasing
function $\chi (t)$,
$$\mu (n) = \int _o^\infty t^n d\chi (t) , $$
for a given moment sequence
$\{\mu (n)\}_{n=o}^\infty$ (the Stieltjes problem
\cite{W,BG}) is very  complicated.  Throughout  the  paper
we shall only consider  absolute  continuous  functions
$\chi (t)$.
Instead of a moment sequence, we  shall  prefer to choose a
suitable weight function. Such a weight  function  could be
$\exp (-\exp t)$. The reasons are that the moment
constant summability method with the moment sequence
$\{\mu (n)\} _{n=o}^\infty$ where $\mu (n)$ are given by (\rf{m3})
plainly  belongs
to the moment  constant  summability  methods  discussed in
\ct{H}, and as such it is regular. The next reason is
yielded by the following Lemma.\vst{0.3cm}\nl

\blm : The function
          \begin{equation}
          \mu (s) := \int _o^\infty \exp (-\exp t)t^s dt
\label{h15}
           \end{equation}
          1) has a meromorphic extension onto the whole complex plane
with simple poles on the negative real axis;\nl
2) its asymptotic behaviour for s tending to infinity,
$\mid\arg s\mid <\pi$, is governed by the saddle point
only and
          \begin{equation}
          \mu (s)\sim [2\pi w(s)w'(s)]^{1/2} \exp [-\exp w(s) +
          s\ln w(s)].
\label{h18}
          \end{equation}
The saddle point $w(s)$ is given in the complex plane by the equation
          \begin{equation}
          (\exp t)t = s    .
\label{h17a}
          \end{equation}
          3) The function $\mu (s)$ exhibits no zero for $Re\ s>-1$.
\lb{rlm1}
\elm
Before the proof we give some comments on the solutions of (\rf{h17a})
for $s$ complex. In this case we shall always assume
that the $s$-complex plane is cut along either the real
positive either the real
negative axis. Correspondingly a sheet in the
$w$-complex plane is chosen  which is  bounded either by the real
positive axis from below and by a curve \(\phi +\arctan (\phi/w) =
2\pi \)
when $Rew\geq 0$, \(\phi +\arctan (\phi/w) = \pi \) when $Rew\leq 0$,
from above, $\phi $ being $Im w$,
$0\leq Imw< 2\pi$. ; either by the curves \(\phi +\arctan (\phi/w) =
\pi\) when
$Rew\geq 0$, $\phi\stackrel{\textstyle>}{<}0$, and \(\phi +\arctan
(\phi/w)=0\)
when $Rew\leq0$, $\phi\stackrel{\textstyle >}{<}0$, $-\pi<Im w<\pi$.
\vs{0.5cm}\nl
{\em Proof}  : 1) The integral on the r.h.s. of (\rf{h15}) converges
absolutely in the whole complex halfplane $Re s>-1$  and
defines there  an  analytic function. To perform an analytic
continuation we make use of the standard Cauchy procedure.
Being
an entire function, $\exp (-\exp t)$   possesses a power
series expansion,
$$\exp (-\exp t) = \sum _{n=o}^\infty  c_n t^n,$$
with infinite radius of convergence. Then splitting the
integration in (\rf{h15}) into two intervals, one obtains
\begin{equation}
\mu (s) = \sum _{n=o}^\infty \frac{c_n t_o^{n+s+1}}{(n+s+1)} +
\int _{t_o}^\infty \exp (-\exp t)t^s\ dt,
\lb{r9}
\end{equation}
where  the  r.h.s.  of (\rf{r9}) is defined for all $s$ except
for $s=-n-1$,  thereby  providing an analytic continuation of
the integral on the r.h.s. of (\rf{h15}) onto the whole complex plane after
one sets $t_o=1$.

2)  For $s$ tending to infinity and
$\mid\arg s\mid\leq\pi /2$ the asymptotic behaviour of
$\mu (s)$ can be calculated directly using the
saddle  point  technique  on  the integral on the r.h.s. of
(\rf{h15}).
The contribution $V_s$ of the saddle point (\rf{h17a}) is
calculated in the standard manner \ct{F,J}. One rewrites integrand of
(\ref{h15})  as an exponential function $\exp h(t)$, where
\be
h(t) := -e^t +s\ln t,
\lb{uf1}
\ee
and looks for critical points of $h(t)$. This means that one
solves the equation
\be
h'(t) = -e^t + s/t = 0.
\lb{uf2}
\ee
Any solution of (\ref{uf2}) is called the saddle point and its
contribution to (\ref{h15}) is given by the formula
\be
V_s = \left(\frac{-2\pi}{h''(t_s)}\right)^{1/2} \, e^{h(t_s)},
\lb{uf3}
\ee
where $t_s$ denotes the value of a given saddle point. Under some
additional conditions one can prove that for $s\rightarrow\infty$
the value of integral (\ref{h15}) is completely determined by
the contributions of saddle points \ct{F,J}. In our case one finds
that (\ref{uf2}) has an unique solution in the cut complex plane,
which we shall denote by $w(s))$. Since $w(s)\exp w(s) =s$ one finds
that
\be
w'(s) = \frac{1}{e^{w(s)}(w(s)+1)},\hs{1cm}
w(s)w'(s)=-\frac{1}{h''(w(s))}\,\cdot
\lb{uf4}
\ee
Thus the contribution of the saddle point $w(s)$ is given by
$$V_s = (2\pi w(s)w'(s))^{1/2} \exp(-e^{w(s)} +w(s)\ln s)$$
and  the asymptotic behaviour of $\mu (s)$ is completely
governed by it \cite{F,J}, so that
$$\mu (s)\sim (2\pi w(s)w'(s))^{1/2}[\exp (-e^{w(s)} +w(s)\ln s)]. $$
For  $s\rightarrow\infty$ and $\mid\arg s\mid >\pi /2$ the
situation is quite different, because  the contribution
$V_e$ of the end point of
the integral on the r.h.s. of (\rf{r9}) \cite{F,J}
$$ V_e = -t_o^s \exp (-\exp t_o )\, /s\hspace{1cm}
(s\rightarrow\infty ),$$
is  far  from  being negligible and in fact is greater than
the  contribution $V_s$ of the saddle point which
 approaches  zero
when $s$ tends to infinity, and
$\pi /2<\mid\arg s\mid <\pi$.
However, we can prove that the contribution $V_e$  is
nothing but the
sum on the r.h.s. of (\rf{r9}) in the large $s$  limit with
opposite
sign, so that these contributions cancel each  other and
only the contribution of the saddle point $V_s$  survives.
In fact,
\begin{equation}
\sum _{n=o}^\infty c_n \fr{t_o^{n+s+1}}{n+s+1}
= (t_o^{s+1}/s)\sum _{n=o}^\infty c_n \fr{t_o^n}
{1+(n+1)/s}\, \cdot
\lb{r10}
\end{equation}
The  sum on the r.h.s. of (\rf{r10}) converges uniformly in
$s$, $0\leq\mid\arg s\mid\leq\pi -\varepsilon$,
$$\sum _{n=o}^\infty c_n \fr{t_o^n}
{1+(n+1)/s}\leq (1/\delta ) \sum _{n=o}^\infty c_nt_o^n <
\infty, $$
where $\delta$ is some constant whish does not depend on
$s$, so we have
$$\lim _{s\rightarrow\infty} \sum _{n=o}^\infty
c_n t_o^n
/[1+(n+1)/s] = \sum _{n=o}^\infty c_nt_o^n = \exp(-\exp
t_o), $$
and
$$\sum _{n=o}^\infty c_nt_o^{n+s+1} / (n+s+1)\sim -V_e
\hspace{1cm}(s\rightarrow\infty ).$$
To prove 3) suppose that for some $s_o$,
          \[\mu (s_o)=\int _{\infty}^\infty \exp [-\exp (\exp t) +
          (s_o +1)t] dt = 0 \ .\]
          However, this implies that for any $\delta$,
          \[\int _{\infty}^\infty \exp\{-\exp [\exp (t+\delta)] +
          (s_o +1)t\}\ dt = 0\ ,\]
          what means on the other hand that
\[ \int _\infty ^\infty \exp [-\exp (\exp t) +
(s_o +2)t][-\exp (\exp t)]\ dt = 0 . \]
          Repeating this procedure $m$-times one gets that
          \[I(m) = \int _o^\infty \exp [-\exp t + mt + (s_o+m)\ln t]
          A^{-m}[A(-1 + d/dA)]^{m-1}(-A)\ dt = 0\ ,\]
          where  $A=\exp t$.    Note    that    the  term $D(t)$,
          \[D(t):=A^{-m}[A(-1+d/dA)]^{m-1}(-A) ,\]
          is  bounded on $t\in (0,\infty )$ and nonzero for
          sufficiently  large  $t$.  On  the other side,
provided $m$ is
          sufficiently large we may use an asymptotic formula to
          evaluate $I(m)$. Thus,
          \[I(m)\sim [2\pi
          u'u/(u+1)]^{1/2}\exp [-\exp u+mu+(s_o +m)\ln u]D(u) =
          0,\]
          since $u(m)$ is the solution of the equation
          \[-e^t + m + (s_o+m)/t = 0,\]
          which behaves like
\[u(m)\sim\ln [m + (s_o+m)/\ln m] \hspace{1in}
          (m\rightarrow\infty).\]
$\Diamond$\vst{0.3cm}\nl

If one uses the approximate solution to the saddle point equation
(\rf{h17a})
          \begin{equation}
w(s) = \ln (s/\ln s) + \ln\ln s/\ln s + {\cal O}[(\ln\ln s/\ln s)^2]
           \hspace{1.7cm} (s\rightarrow\infty),
\label{h17b}
          \end{equation}
then one finds the following approximate asymptotic behaviour of
$\mu (s)$,
          \begin{equation}
          \mu (s)\sim (2\pi\ln s/s)^{1/2} [\exp -(s/\ln s)]
          [\ln (s/\ln s)]^s  \hspace{2cm}    (s\rightarrow\infty)\ ,
\label{h16}
          \end{equation}

Some words about a numerical implementation of the method.
Note that the method can be implemented {\em numerically} as
well, since the solution to (\ref{h17a}) is very quickly determined
by the following recursive formula:
\begin{equation}
w(n)=ln(n/ln(n/ln(n/\ldots))),
\end{equation}
and the asymptotic formula (\ref{h18}) very accurately reproduces
the value of the integral (\ref{h15}) (in fact, up to ten orders by
a very simple algorithm).
One also  finds that and when $s$ tends to infinity and
$|\arg s| < \pi$ the relation (\ref{uf4}),
\(w'(s)\sim 1/s-1/sw(s)\), still holds.\vst{0.3cm}\nl
{\bf\em Remark} 1 : Note that the proof of Lemma \rf{rlm1} can be directly
adapted to the case of the weight function $\exp[-e_k(t)]$,
where $e_k (t)$ is the $k$-fold exponential function,
$$e_k (t) := \exp(\exp(...(\exp t)...),$$
$k$ being an arbitrary
non-negative integer. The saddle point is now given by the
equation
$$e_k (t) e_{k-1}(t) ... e^t t = s.$$
In the case of $k=0$ one obtains a simpler proof of the
asymptotic
properties of the gamma function than the one usually known
\cite{E}. For $k\geq 2$ one obtains another moment
constant summability method
$\mu _k$ solving the problem ({\bf B}) (see
below). They give nothing new in the regular case, but they
play quite an important role in the singular case.
Considering the integral on the r.h.s. of (\rf{h15}) as the Mellin
transform, one could adapt Lemma {\rf{rlm1} to study the Mellin
transform of some class of entire functions. $\diamondsuit$\vst{0.3cm}\nl

Now, our main problem is the determination of
asymptotic properties of the moment function $F(t)$, defined by
\begin{equation}
F(t) :=  \sum _{n=o}^\infty\frac{t^n}{\mu (n)}\,\cdot
\lb{r11}
\end{equation}
The reason is that for $\mid z\mid < 1$
\begin{equation}
\int _o^\infty \exp(-\exp t)\, F(tz)\, dt\, =\, \sum _{n=o}^\infty
z^n\, =\, 1/(1-z) ,
\lb{r12}
\end{equation}
due  to  the  regularity  of the $\mu$-method
\cite{H}. Because of
Lemma \rf{rlm1}  the  function $F(t)$ is an entire function. One
expects that
$$\max _{\mid t\mid =const} \ln\mid F(t)\mid\sim\exp
(\mb{const} \exp\mid t\mid )\hspace{1cm} (t\rightarrow\infty )$$
\cite{J,L}.
If  $F(t)$ grew no faster than $\exp\mid t\mid ^A$ when $t$
approaches infinity, and
 \mb{$\mid\arg t\mid >0$}, where $A$ is a  constant,  then  (\rf{r12})
should converge absolutely for every
$z \in MLS[1/(1-z)]$, the Mittag-Leffler star of the Cauchy
kernel $1/(1-z)$ providing
in such way an analytic continuation of
$\sum _{n=o}^\infty z^n$ from the unit disc onto the whole
$MLS[1/(1-z)]$.

Let $f(z)$ be the
principal branch of an analytic function regular at origin.
Then
\[
f(z) = (1/2\pi i)\oint _C \frac{f(u)}{u(1-z/u)} \ du
\hs{5cm}\]
\be
= (1/2\pi i)\oint _C \frac{f(u)}{u}\ du\int _o^\infty
          \exp(-\exp t) F(tz/u)\ dt
\lb{r13}
\ee
for every simple contour $C$ such that no singularity of
$f(z)$ lies on $C$ or inside it. If in addition  the  contour
$C$ is such that
$\{z/u\mid u\in C\}\subset MLS[1/(1-z)]$, then  we  can
invert the order of the integrations in (\rf{r13}), to obtain
\bea
\lft{f(z) =\int _o^\infty \exp(-\exp t)
\sum _{n=o}^\infty\frac{(tz)^n}{\mu (n)}\ \left(\frac{1}{2\pi i}\right)
\oint _C \frac{f(u)}{u^{n+1}}\, du\, dt}\hs{5cm}\nonumber\\
& & = \int _o^\infty \exp(-\exp t)
          \sum _{n=o}^\infty a_n\frac{(tz)^n}{\mu (n)}\, dt  ,
\lb{r14}
\eea
$a_n$ being the Taylor coefficients of $f(z)$ at origin. It
is a short exercise to show that the set  of
$z$  for  which  the $\mu$-sum (\rf{r14}) exists is just $MLS(f)$.
Hence  to  show that the
$\mu$-method provides an analytic continuation of the
Taylor series of $f(z)$ at origin onto the whole $MLS(f)$
we have
to prove that $F(t)$, as defined by (\rf{r11}), does not  grow
faster than $\exp(\mid t\mid ^A)$ when $t$ tends to infinity
and $\mid\arg\ t\mid >0$.
One can also deal with the case where some finite
number of singularities of $f(z)$, but no branch point, are
elements of $C_o$ (interior
domain with respect to $C$) provided the condition
$\{z/u \mid u\in C\}\subset MLS[1/(1-z)]$ is satisfied. The
result is
\begin{equation}
f(z) = -\sum '_s Res[f(u),z_s]/(z_s -z)
+\int _o^\infty \exp(-\exp t) \sum _{n=o}^\infty
          \tilde{a}_n \frac{(zt)^n}{\mu (n)} dt ,
\lb{r15}
          \end{equation}
          where
          \begin{equation}
          \tilde{a}_n
          = a_n + \sum '_s Res[f(u),z_s]/z_s^{n+1} ,
\lb{r16}
          \end{equation}
and $\sum '_s$ in (\rf{r15},\rf{r16}) runs over all singularities
$z_s$ of
$f(z)$, which are elements of $C_o$ (see also the next
section). As an exercise one can justify the relations
(\rf{r15},\rf{r16}) for the Cauchy kernel
$1/(1-z)$, where $\tilde{a}_n =0$ so that the integral on  the
r.h.s. of (\rf{r15}) vanishes.

Let  us turn now to the study of asymptotic properties
of the function $F(t)$.\vst{0.3cm}\nl
\blm Let
$\{\mu (n)\}_{n=o}^\infty$ be the Stieltjes moment
sequence generated by the measure
$\exp (-\exp t)\ dt$ (\rf{m3}). Then
                    the function $F(t)$, defined by (\rf{r11}),
$$F(t) = \sum _{n=0}^\infty\frac{t^n}{\mu (n)},$$
is an entire function with the following
asymptotic behaviour at infinity :\nl
1) For
$\mid Im t\mid \leq\pi /2$ the asymptotic behaviour is
determined by the saddle point
$s = (\exp t)t$, of the Euler-Maclaurin integral representation
of F(z),
$$
F(t) = \int_{\sigma}^\infty \frac{e^{s\ln t}}{\mu (s)}\ ds +
{\cal O}(\mid t\mid^\sigma )\hspace{2cm}
(t\rightarrow\infty ),
$$
$\sigma$  being some constant, $-1<\sigma <0.$
Therefore
$$F(t)\sim \exp [\exp t +t +\ln(t+1)]\hspace{1.5cm}
(t\rightarrow\infty ).$$
2) For $\mid Imt\mid >\pi /2$ the asymptotic behaviour of F(z) is still
determined by the saddle point whenever its contribution prevails
the contribution of the end point of integration. Otherwise, as well as
for $|Im\,t|>\pi$,
\[\mid F(t)\mid\leq  {\cal O}(\mid t\mid^\sigma )
\hspace{1in} (t\rightarrow\infty ).\]
\lb{rlm2}
\elm
{\em Proof}  : By Lemma \rf{rlm1} $\mu (s)$  exhibits  no  zero in the
right  complex halfplane. Then by virtue of the Euler-Maclaurin sum
formula \cite{F,J} we arrive at the ensuing
integral representation of $F(t)$,
\begin{equation}
F(t) = \int_{\sigma}^\infty \frac{e^{s\ln t}}{\mu (s)}\ ds +
{\cal O}(\mid t\mid^\sigma )\hspace{1cm}
(t\rightarrow\infty ),
\lb{r17}
\end{equation}
where $\sigma$ is some noninteger from the interval $(-1,0)$.
To find an asymptotic behaviour of $F(t)$ at infinity we shall use the
standard saddle point method as in the proof of Lemma 1. In the present
case one does not know an explicite form of the function $h(s)$ (\ref{uf1})
which determines the position of saddle points.
Nevertheless the position of
the saddle point of (\ref{r17}) can be found exactly for sufficiently
large $t$.
As one expects and as we shall show in a moment, function $h(s)$, and
thus  equation for the saddle point is
determined by the asymptotic behaviour of
$\mu (s)$, provided $t$ is sufficiently large. Let as above
$w=w(s)$ be the solution of (\rf{h17a}), i.e.,
\begin{equation}
w(s)\exp [w(s)] = s .
\lb{r18}
\end{equation}
Due to Lemma \rf{rlm1} the asymptotic behaviour of $\mu (s)$ at infinity
is
\begin{equation}
\mu (s)\sim [2\pi w(s)w'(s)]^{1/2}
          \exp\{-\exp [w(s)]+s\ln w(s)\}.
\lb{r19}
          \end{equation}
This in turn determines an asymptotic form
of the function $h(s)$,
          \begin{equation}
          h(s) = \exp [w(s)]-s\ln w (s)+s\ln t   ,
\lb{r20}
          \end{equation}
and hence a solution of the equation
 \be
          h'(s) = e^w w'(s)-\ln w-sw'(s)/w(s)+\ln t = 0 .
\ee
By using the relation (\rf{h17a}) the last equation is essentially
reduced to
          \begin{equation}
          h'(s) = -\ln w+\ln t = 0,
\lb{r21}
\end{equation}
and the saddle point is $t=w(s)$. From (\rf{r21}) one finds that
$h''(s)=-w'(s)/w(s)$ at the saddle point. Finally, from
(\ref{uf3}) one obtaines
the contribution $V_s$ of the saddle point,
\be
V_s := \left(\frac{2\pi w(s)}{w'(s)}\cdot\frac{1}{2\pi w(s)w'(s)}\right)^{1/2}
e^{e^t}
= [w'(s(t))]^{-1} e^{e^t} := e^{\omega (t)} ,
\ee
where
\be
\omega (t):= \exp t+t+\ln (t+1).
\lb{om}
\ee
If $\mid Im t\mid\leq\pi /2$  then one can justifies
that the asymptotic behaviour of $F(t)$ is determined by the
contribution of the saddle point only.
If $\pi /2<\mid Im t\mid <\pi$ then the asymptotic behaviour is still
determined by the
contribution of  the  saddle  point  provided it prevails the
contribution of the end point of the integral (\rf{r17}).
For example this happens
if one stays on special curves which approximates $y=\pi/2$
($y=-\pi/2$) from above (below) if $x\rightarrow\infty$ (see also
below section \rf{sub:sing}).
Otherwise, and for $|Im\,t|>\pi$ as well, the contribution of the end
point $s=\sigma$ of the integral (\rf{r17})  prevails and $|F(z)|$
tends to zero.
If $\mid Im t\mid\geq\pi$ the contour of the
steepest descent does not exist since the saddle point
does not lie  on  the  first  sheet of the multivalued
function $w(s)$.
Another point of view is that we do not have any saddle point
of the integral (\rf{r17}) since $w(s)$ in (\rf{r19}) is only defined for
$|\arg s\mid <\pi$ for which $\mid Im\, w(s)\mid <\pi$. Hence,
the relation (\rf{r21}) cannot be satisfied for such $t$.
In both cases, however, one
can show that the  asymptotic behaviour of
$F(t)$ is determined by the end point
$s=\sigma$  of  the  integral  (\rf{r17}),  i.e.,
$F(t)\sim{\cal O}(\mid t\mid^\sigma )$ for $t$  tending  to
infinity and $\mid Im t\mid\geq\pi$ \ct{F,J}. For
$\mid\arg t\mid >0$ this can be examplified
independently as  follows.  By  virtue of the analytic properties  of
$\mu (s)$ and its behaviour at infinity (\rf{h16}) one can rotate
the contour of integration  from  the  real  axis  to  the
ray $(\sigma ,\sigma +iq\infty )$ , where
$q=sign\ (\arg t)$.  Let $\arg t>0$. Then the integral $I$,
\[ I = \int _{\sigma}^{\sigma +i\infty} \frac{e^{s\ln t}}{\mu (s)}\ ds,
\]
can be majorized in the subsequent manner,
    \begin{equation}
          \mid I\mid\leq e^{\sigma\ln\mid t\mid}
          \int _{\sigma}^{\sigma +i\infty} \frac{e^{-y\arg\ t}}
      {\mid\mu (s)\mid}\
          \mid ds\mid\leq 2\mb{const}\ e^{\sigma\ln\mid t\mid},
\lb{r22a}
          \end{equation}
          where $s =\sigma +iy$, because on the ray
          $(\sigma ,\sigma +i\infty )$
          \begin{equation}
\mid 1/\mu (s)|\sim\exp [\mb{const}\ (y/\ln y)]\hspace{2cm}
          (y\rightarrow\infty ).
\lb{r22b}
          \end{equation}
Thus  $I$  decays at worst like $\mid t\mid ^\sigma$ when $t$ tends
to infinity
and the proof is finished. $\Diamond$ \vst{0.3cm}\nl
{\bf\em Remark} 2 : Note that the position of the saddle point is
          $$w(s) = t$$
          also  in  the  case of Stieltjes moment sequence
          $\{\mu _k (n)\}_{n=o}^\infty$
          generated  by $\exp [e_k (t)]$, where $k$ is an arbitrary
          nonnegative integer. Now,
          $w(s)$ is the solution of (\rf{r18}) with
          $\exp w$
          replaced by $(d/dw)e_k (w)$ (see Remark 1). The
          contribution of the saddle point is up to the Gaussian
integral around the saddle point
$$V_s =\exp [e_k (w)]   \hspace{2cm}
          (t\rightarrow\infty ).$$
          Analogously as in the above case of $k=1$ ,
          $F_k (t)$, defined by
          $$F_k (t) = \sum _{n=o}^\infty\frac{t^n}{\mu _k (n)},$$
          where $k\geq 2$, is also polynomially bounded when $t$
          tends to  infinity and $\mid\arg t\mid >0$ because of
          $$\mid 1/\mu _k (s)| \sim\exp\{const
          [y/\ln y...\ln_{k-1}(y)\ln_k (y)]\}\hspace{1cm}
          (y\rightarrow\infty ),$$
on  the  ray $(\sigma ,\sigma +iq\infty )$,
$\ln_k (y)$  being the $k$-fold logarithm. It means that
such  methods also solve  the problem
({\bf B})! The principal difference  between  the  Borel $(k=0)$ and
$\mu _k$-methods with $k\geq 1$ consists in the fact that for  the
Borel method
the  integral  on the r.h.s. of (\rf{r22a}) may not  converge  if
$\mid\arg t\mid >0$. $\diamondsuit$\vst{0.3cm}\nl

          The main result of this section is the following Theorem.
\vst{0.3cm}\nl
\bth : Let $f(z)$ be the principal branch of an analytic function
regular at origin,
 $$f(z) =  \sum _{n=o}^\infty a_n z^n  .$$
                      Then the integral $I(z_o)$,
\be
I(z_o ) =\int _o^\infty \exp (-\exp t)
           \sum _{n=o}^\infty a_n \frac{(z_o t)^n}{
          \mu (n)}\ dt ,
\lb{r23}
\ee
converges if and only if $z_o\in MLS(f)$, where
$\mu (n)$ are defined by (\rf{m3}). If $z_o\in MLS(f)$ then
  $$f(z_o ) = I(z_o ).$$
The convergence is absolute and uniform in any bounded subset of $MLS(f)$
with nonzero distance from the boundary of $MLS(f)$ and we can
differentiate inside the sign of integration,
  $$ f'(z) =\int _o^\infty\exp (-\exp t)
          \sum _{n=1}^\infty na_n \frac{z^{n-1}t^n}{\mu (n)}\
          dt.$$
          Provided $f(z)$ has no branch point on or inside $C$,
then the relations (\rf{r15},\rf{r16}) hold.
\lb{rth1}
\eth
{\em Proof}  : Note that the integrand of (\rf{r23}) is an entire
function, so that it is well  defined. The result follows
immediately from Lemma \rf{rlm2} and the relations
(\rf{r13},\rf{r14}). $\Diamond$\vst{0.3cm}\nl
{\bf\em Remark} 3 : $MLS(f)$ is invariant with respect to
           differentiation, i.e.,
           $MLS[f(z)]\\ = MLS[f'(z)]$. If $f(z)$ is as in Theorem
\rf{rth1},  then $f'(z)$ also satisfies the conditions of this
          theorem, and we have another representation of $f'(z)$,
 $$f'(z) =\int _{o}^\infty\exp (-\exp t)
\sum _{n=1}^\infty n a_n\frac{(zt)^{n-1}}{\mu (n-1)}\ dt,$$
i.e., the moments $\mu (n)$ can be in some sense shifted.
$\diamondsuit$\vst{0.3cm}\nl

Let  $f(z)$ be  as in Theorem \rf{rth1} and regular on the real
positive  axis,  for instance. Then Theorem \rf{rth1} enables us to
          give an analytic  continuation of the Taylor series of
          $f(z)$ at origin on the whole axis. In fact the following
          Corollary holds.\vst{0.3cm}\nl
\bcl : Let

 1) $f(z)$ be the principal branch of an
                           analytic function regular at origin;

 2) $f(z)$ be regular on the real positive axis.

                        Then

 $$f(x) =\int _o^\infty \exp (-\exp t)
           \sum _{n=o}^\infty  a_n \frac{(xt)^n}{\mu (n)}\
          dt$$
for  $x\in [0,\infty )$. The integral converges absolutely
and one can differentiate inside the sign of integration.
\lb{rcl1}
\ecl
\np

\subsection{Some new results on the Borel summability method}
\lb{sub:some}

In  this  section we wish to prove an analogue of Theorem \rf{rth1}
for the Borel method. Some results are exposed in \cite{SG}, but a
detailed  study  of  the region of convergence of the Borel
integral is missing. The same role as the function $F(t)$ has in the
$\mu$-method discussed above plays the generalized
Mittag-Leffler  function in the Borel method. As we have mentioned
above the Borel method
is one member  of  the moment constant summability methods.
Its weight  function is
$(1/\alpha )t^{(\beta/\alpha )-1} \exp (-t^{1/\alpha})$, and
the moments of the Borel method are in general
 $$ (1/\alpha )\int _o^\infty t^{(\beta/\alpha
          )-1} e^{-t} \ t^n\ dt =\Gamma (\alpha n +\beta ) ,$$
where $\Gamma $ is the usual gamma function, and
          $\alpha ,\beta $ are positive constants, $0<\alpha\leq 2$
          (note  that  the  Borel  method  with
          $\alpha =\beta =1$ can be viewed  as a special case
          $\mu _o$ of the $\mu _k$-methods discussed in Remark 1).
          The generalized Mittag-Leffler function is the
          function
 $$E_{\alpha ,\beta}(z) = \sum _{n=o}^\infty \frac{z^n}{\Gamma
          (\alpha n +\beta )}$$
\cite{E}.  We wish to show that the problems ({\bf A}) and
({\bf B}) of the
          Borel  method are results of the fact that
          $E_{\alpha,\beta}(z)$ is unbounded in  some {\em sector-like
 domain}  with {\em nonzero opening
          angle}. Indeed, its asymptotic properties when $z$ approaches
          infinity are as follows \cite{H,E}:

          a) $$E_{\alpha ,\beta} (z) \sim\sum_{n=1}^\infty z^{-n}
          /\Gamma (\beta -\alpha n)\hspace{1cm}
          if\hspace{0.5cm}\mid\arg (-z)\mid < (1 - \alpha /2)\pi ,$$

 b) $$E_{\alpha,\beta} (z) \sim (1/\alpha)\sum_m t_m^{1-\beta}
          e^{t_m} \hspace{1cm} if\hspace{1cm}\mid\arg z\mid\leq
          \pi\alpha /2 ,$$
          where
          $t_m = z^{1/\alpha} e^{2\pi im/\alpha}$ ,
          and  the  sum  runs over all $m$ such that
          $-\pi\alpha /2\leq\arg z + 2\pi m\leq\pi\alpha /2.$
          Because of these asymptotic properties the integral
         \begin{equation}
          \int_o^\infty t^{\beta -1} e^{-t} E_{\alpha ,\beta}
         (zt^\alpha )dt ,
\lb{r24}
         \end{equation}
 which  is  for $\mid z\mid <1$ equal to $1/(1-z)$, {\em does not}
 converge in
          the   whole  $MLS[1/(1-z)]$   but   only   in   some   domain
          ${\cal B}_\alpha [1/(1-z)]$,
          \begin{equation}
          {\cal B}_\alpha [1/(1-z)] := \{ z\mid Re(z)^{1/\alpha} =
          r^{1/\alpha}[\cos (\theta/\alpha)]< 1\}.
\lb{r25}
\end{equation}
Note  that  ${\cal B}_\alpha [1/(1-z)]$  does not depend on
$\beta$. Due to the
relations (\rf{r13},\rf{r14}) such a representation of the Cauchy
          kernel determines the representation  of  an arbitrary function
          $f(z)$ regular at origin. Let $C(z_o )$ be a contour given by
          the relation
          \begin{equation}
          r = r_o\{\cos[(\theta -\theta_o)/\alpha ]\}  \hspace{0.5cm}
\mb{where}\hspace{0.5cm}\mid\theta -\theta_o\mid <\pi\alpha /2.
\lb{r26}
         \end{equation}
Let  us  draw  the contour $C(z_s )$ for each singularity $z_s$ of
$f(z)$  and  let us discard from the complex plane the domain
$C_o (z_s )$  closed  up  by the contour, for which the sign =
in (\rf{r26}) is  replaced by $>$. Then the notation
${\cal B}_\alpha (f)$ is adopted
for the simply  connected  region of the complex plane
containing the origin.  By  the construction
${\cal B}_\alpha (f)$ is a starlike region containing, in any
case, the disc of convergence of the Taylor series of $f(z)$
around the origin.\vst{0.3cm}\nl

\bth : Let $f(z)$ be the principal branch of an analytic
                      function, regular at origin,
 $$f(z) = \sum_{n=o}^\infty  a_n z^n .$$
Then the integral $I(z_o )$,
\begin{equation}
          I(z_o ) = \int_o^\infty  t^{\beta -1}e^{-t}
          \sum_{n=o}^\infty a_n
\frac{(z_o t^\al )^n}{\Gamma(\beta +\alpha n)}\ dt,
\lb{r27}
          \end{equation}
  converges if and only if $z_o \in{\cal B}_\alpha (f)$.
If $z_o\in {\cal B}_\alpha (f)$ then
                    $$f(z_o ) = I(z_o ) .$$
The convergence is absolute and uniform on any bounded
subset of ${\cal B}_\alpha (f)$ with nonzero distance from
the boundary of ${\cal B}_\alpha (f)$. For the first
derivative $f'(z)$ of $f(z)$  the following representation
\begin{equation}
f'(z) = \int_o^\infty  t^{\beta -1} e^{-t}
\sum_{n=o}^\infty na_n\,
\frac{z^{n-1} t^{\alpha n}}{\Gamma(\beta +
\alpha n)}\ dt
\lb{r28}
\end{equation}
holds.
\lb{rth2}
\eth
{\em Proof}  :  Proof of Theorem is nothing but a slight
modification of the proof of Theorem \rf{rth1} to the case  of
another Stieltjes moment sequence. The double integral on the
r.h.s. of (\rf{r13}) converges if and only if  for  a  given
$z_o$ a contour $C$ exists such that
$\{ z_o /u \mid u\in C\}\subset {\cal B}_\alpha [1/(1-z)] .$
Interchanching  the  order  of integrations in (\rf{r13}) one
obtains the relation (\rf{r27}) if and only if:

a) the origin and $z_o$ are elements of $C_o$ ;

b) no singularity $z_s$ of $f(z)$ is an element of $C_o\cup C$;

c) $\forall u\in  C$, where $u =\rho\exp (i\phi )$ and such
          that $\mid\theta_o -\phi\mid <\pi\alpha /2$, the inequality
          \begin{equation}
          Re\ (z_o /u)^{1/\alpha} = (r_o /\rho)^{1/\alpha} \cos
         [\theta_ o -\phi )/\alpha ]\leq 1-\delta
\lb{r29}
          \end{equation}
          holds, where $\delta$ is a positive constant.

The  first two conditions are obvious; the last one follows
from the fact that the integral (\rf{r24}) converges if and  only
          if
 $$Re\ z^{1/\alpha} \leq 1 -\delta .$$
          One  can easily check that a set of all $z_o$ such that a
          contour  $C$ with the above properties exists is just the
          region ${\cal B}_\alpha (f)$.  In  fact,  for all
          $z\in{\cal B}_\alpha (f)$ such a contour
          exists. If $z\not\in {\cal B}_\alpha (f)$, then
          inevitably a singularity $z_s$ of $f(z)$ exists such that
$$r\geq r_s\{\cos [(\theta -\theta_s )/\alpha]\}^{-\alpha}
 \hspace{0.5cm}\mb{when}\hspace{0.5cm}\mid\theta -\theta_s\mid
         <\pi\alpha /2.$$
This, however, contradicts to the properties b) and c)
          under which
$$r_s > r\{\cos [(\theta -\theta_s )/\alpha ]\}^{\alpha} >
 \rho \hspace{0,5cm}\mb{if}\hspace{0.5cm}\mid\theta -\theta_s\mid
          <\pi\alpha /2.$$
  The relation (\rf{r28}) then follows immediately from the
          absolute convergence and the fact that under  the  conditions
of the  Theorem the integrand of (\rf{r27}) is an  entire  function.
$\Diamond$\vst{0.3cm}\nl
{\bf\em Remark}  4 : Like $MLS(f)$ the domain ${\cal B}_\alpha (f)$
is also invariant under  differentiation,
          ${\cal B}_\alpha (f') = {\cal B}_\alpha (f)$, so that
          inserting the first derivative $f'(z)$ in place of $f(z)$ in
Theorem \rf{rth2} one
obtains another integral representation  of  $f'(z)$,  like  in
the previous section,
$$f'(z) = \int_o^\infty t^{\beta -1} e^{-t} \sum_{n=1}^\infty
na_n\, \frac{(zt^\alpha )^{n-1}}{\Gamma [\beta +\alpha (n-1)]}\ dt.\hs{1cm}
\diamondsuit\vst{0.3cm}\nl$$
{\bf\em Remark}  5 : Note   that   whenever $\alpha <\alpha '$, then
          ${\cal B}_\alpha (f)\subset{\cal B}_{\alpha '} (f)$. For
          $\alpha\rightarrow 0$ the domain ${\cal B}_\alpha (f)$
          approaches $MLS(f)$. However, the limit
$$ \lim_{\alpha\rightarrow 0}\Gamma (\beta +\alpha n)
  = \Gamma (\beta )$$
is trivial (does not depend on $n$). This fact provides one of arguments
for the moment sequence of the moment summability methods summing the
Taylor  series  of  $f(z)$  in  the whole $MLS(f)$ to grow
like $(\ln n)^n$, because
 $$\mu (n)\sim (\ln n)^n= o\,[\Gamma (\beta +\alpha n)]
          \hspace{1cm} (n\rightarrow\infty),$$
for  all positive $\alpha$ and $\beta$. Another way is to
consider the weight function for the Borel moments which
contains a factor $\exp (-t^{1/\alpha})$ and its behaviour
when $\alpha\rightarrow 0_+$. $\diamondsuit$\vst{0.3cm}\nl

 Theorem  \rf{rth2}  shows  that  in the regular case the Borel
          summation is not only the Laplace transform (cf.
          \cite{RS}). The integral (\rf{r27}) in the Laplace form,
 $$I'(z_o ) = (1/z_o)^{1/\alpha} \int_o^\infty
  t^{\beta -1} \exp (-t/z_o^{1/\alpha}) \sum_{n=o}^\infty
 a_n \frac{t^{\alpha n}}{\Gamma (\beta +\alpha n)}\ dt,$$
has  a different region of convergence but one is an
analytic continuation of the other. In fact, adopt the
notation $L(z)$ ($L^o (z)$) for the simple contour
(the interior domain with respect to it), where $L(z)$ is
parametrized by $u := \rho\exp (i\phi )$ in the following way
(see Fig. \rf{fg4}):

          a) If $ \mid\theta -\phi\mid\leq\mid\theta -\phi_o\mid
          <\pi\alpha /2,$ where $z = r\exp (i\theta )$, then
          \begin{equation}
          \rho = r\{\cos [(\theta -\phi )/\alpha ]\} ,
\lb{r30}
          \end{equation}
 $\pm\phi_o$ being determined by the equation
  $$\rho = r\, [(\cos (\theta -\phi )/\alpha ]^{\alpha} =
          \varepsilon ,$$
 where $\varepsilon$ is strictly less than radius of convergence of the
          Taylor series at the origin of the function under
          consideration;

          b) If $\mid\theta -\phi\mid >\mid\theta -\phi _o\mid$ ,
  $\pm\phi _o$ being the same as above, then
          \begin{equation}
          \rho = \varepsilon .
\lb{r31}
          \end{equation}

          The following theorem holds.\vst{0.3cm}\nl

\bth : Let $f(z)$ be the principal branch of an analytic
                      function regular at origin. Denote by
                      $D_{\alpha} (f)$ the domain
          \begin{equation}
           D_{\alpha} (f) := \cup _{z\in{\cal B}_\alpha (f)}
          L^o (z),
\lb{r32}
\end{equation}
where $L^o (z)$ is defined by (\rf{r30},\rf{r31}).
Thus for any $z\in D_{\alpha} (f)$, $z=r\exp (i\theta )$, there
exists  $z^*=r^*\exp (i\theta^*)\in{\cal B}_\alpha (f)$
such that $z\in L^o (z^*)$.
Let $z_o$ be
$$z_o := z \exp(-i\theta^* ). $$

Then the
          integral
          \begin{equation}
          I'(z) :=  (1/z_o)^{1/\alpha} \int_o^\infty  t^{\beta -1}
          \exp(-t/z_o ^{1/\alpha}) a_n (e^{i\theta^ *} t^\alpha )^n
          /\Gamma  (\beta +\alpha n)\  dt
\lb{r33}
          \end{equation}
          converges (outside the Borel polygon!) and equals to $f(z)$ .
\lb{rth3}
\eth
{\em Proof} : Suppose the corresponding
          $z^* \in{\cal B}_\alpha (f)$ has been found. Then the integral
 $$\int_o^\infty t^{\beta -1} e^{-t} \sum_{n=o}^\infty
 a_n\frac{(z^* t^\alpha )^n}{\Gamma (\beta +\alpha n)}\ dt$$
          converges.  After  the  real substitution $t\rightarrow
          tr^{1/\alpha}$ one gets
          the integral in the Laplace form,
 $$(1/r)^{1/\alpha}\int_o^\infty t^{\beta -1} \exp
 (-t/r^{1/\alpha}) \sum_{n=o}^\infty a_n \frac{(e^{i\theta^*}
          t^\alpha )^n}{\Gamma (\beta +\alpha n)}\ dt,$$
          which  converges $\forall w\in C_o$ (the complex plane of
          $r$) such that
$$Re\, w^{1/\alpha} > r^{1/\alpha},$$
or,  what  is  the same,
$$w<r[\cos (\theta/\alpha )]^\alpha.$$
But the complex plane $C_o$ is nothing but the rotated $C$,
i.e., $C_o = \exp (-i\theta^* )C$. $\Diamond$\vst{0.3cm}\nl

          Let us give an example to illustrate the Theorem.
          Consider the function $f(z) = 1/(1-z)$ and  the  standard
          Borel method ($\alpha =\beta = 1$). Now, ${\cal B}_1 (f)$ is
          the complex halfplane $Re z<1$, and
          $D_1 (f)\equiv MLS(f)$.  Let us calculate, e.g., $f(2+3i)$.
          As the corresponding  $z^*$  we choose $z^* = 8i$. Hence
$\theta^* =\pi /2$ so we have $z_o  =  3-2i$  (see Fig. \rf{fg4}).
\bfg
\vs{8cm}
\caption{{\em Analytic continuation beyond $Re\, z<1$ .}}
\lb{fg4}
\efg
According to (\rf{r33})
\beas
\lft{f(2+3i) = (3-2i)^{-1} \int_o^\infty \exp [-t/(3-2i)]
 \sum_{n=o}^\infty \frac{(it)^n}{n!}\ dt} \\
 & &= (3-2i)^{-1}
          \int_o^\infty \exp\left\{ -t\left(\fr{1}{3-2i} - i\right)
\right\}\ dt = -\fr{1}{1+3i} \cdot
\eeas
\vs{0.5cm}\nl

Theorem \rf{rth3} shows the principal obstruction for the
          Borel  method  to be convergent in the whole $MLS$ of
          function under  consideration.  It  is due to the fact that
          whenever the Borel sum  exists  for some $z_o$, then it also exists
   for $z\in L^o (z_o)$ and defines an analytic function
          there.

          At  the  end  of this section we wish to show that the
          Borel  method  can represent a function regular on the real
          positive  axis only under assumption of its regularity in a
          larger sector-like domain.\vst{0.3cm}\nl

\bcl : Let $f(z)$ be regular in a sector-like domain
                        $S_{\gamma}$ ,
  $S_{\gamma} := \{z\mid r>0\  and\  \mid\theta\mid <\gamma\}.$
Then the integral
 $$(1/z)^{1/\alpha}\int_o^\infty t^{\beta -1} \exp
 [-t/z^{1/\alpha} ] \sum_{n=o}^\infty a_n \frac{t^{\alpha n}}
{\Gamma (\beta +\alpha n)}\ dt$$
 converges  for $z\in (0,\infty )$ if and only if
$\alpha <2\gamma /\pi$. If $\alpha <2\gamma /\pi$, then the
integral converges on each sector $S_{\eta}$,
$S_{\eta} := \{z\mid  r>0 \ and\  \mid\theta\mid <\eta\},$
where $\eta <\pi\alpha /2$.
\lb{rcl2}
\ecl
{\em Proof}  : Only under the condition $\alpha <2\gamma /\alpha$
the real positive axis belongs to ${\cal B}_\alpha (f)$.
$\diamondsuit$\vst{0.3cm}\nl

Note  that all results of the section can be generalized to
the case where some singularities of $f(z)$, but  no
branch point, lie in $C_o$.
Note also that  for  the class of functions satisfying the
hypotheses of the Nevanlinna theorem the Borel transform is
analytic in a strip  including  the real positive axis. So,
applying Corollary \rf{rcl1}  one  directly finds an expression
for the Borel transform on the whole axis (in terms of $a_n'{}^s$).

As  for  the  generalization  of the strong asymptotic
condition one can see  that  by  replacing the factor $N!$ in
the above bound of $R_N (z)$  by  $\mu _k (N)$ one obtains
uniqueness
theorems  for  horn-shaped  regions, $k$  characterizing  the
sharpness of a horn. The proof is simply performed by
combining  conformal
mappings with the Phragmen-Lindel\"{o}f theorems \cite{M2}.\np

\subsection{Singular case}
\lb{sub:sing}

As it has been shown above the moment constant summability method
with the moment sequence $\{\mu (n)\} _o^\infty$,
          \begin{equation}
              \mu (n) := \int _o^\infty \exp (-\exp t) t^n dt .
          \end{equation}
solves the shortcoming ({\bf B}) of the Borel method.
In this subsection we shall show that this method can
be used for the horn-shaped singularity as well.  We  shall
only deal with the horn $H_R$ (see Fig. \rf{fg5}), defined as
\begin{equation}
H_R := \{z\mid Re\ \omega (1/z)>\omega (1/R)\},
\label{h6a}
\end{equation}
where
\begin{equation}
\omega (z) := \ln F(z),\hspace{2cm} F(z) := \sum _{n=o}^\infty
\fr{z^n}{\mu (n)},\label{h6b}
\end{equation}
i.e.,  roughly  speaking, with the region of the asymptotic
type  $(1,1)$
\cite{S,N,So}.
\bfg
\vs{8cm}
\caption{{\em The  horn $H_R$ (a) and its image $H_R^{-1}$ (b)
under the mapping $z\rightarrow 1/z$.}}
\lb{fg5}
\efg

This can be confirmed as follows. By Lemma \rf{rlm2} we know that
$Re\,\om (z)$ tends to $-\infty$ for $z\rightarrow\infty$ and
$Im\,z>\pi$. On the other hand, within the strip $Im\, z\leq\pi$,
$\om (z)$ behaves according to (\rf{om})
as $\om (z)=e^z +z+\ln(z+1)$
provided $Re\,\om(z)$ does not decrease too fast. Thus within the strip
the equation $Re\,\om (z)=const$ which defines the {\em separatrix}
of the asymptotic behaviour of $F(z)$ is nothing but
\be
Re\,\om (z)=e^x\cos y +x+\ln|x+iy+1|=c,
\lb{oi}
\ee
$c$ being some real constant. The Eq. (\ref{oi}) shows that
$\partial H^{-1}_R$ is symmetric under $y\rightarrow\pm y$.
Hence we can restrict our considerations to the right upper quadrant.
In order that $Re\,\om (z)$ be a constant the exponential term in
(\rf{oi}) forces $y$ to move very rapidly in the region $\pi/2<|y|<\pi$
and then to approach very close $|y_o|=\pi/2$ from above if $x$ tends
to infinity.
One anticipates
that the boundary $\partial H_R$ of $H_R$ encloses in some sense the
strip $Im\, z\leq\pi/2$. Indeed, asymptotically, for
$x\rightarrow\infty$, the boundary of $\partial H^{-1}_R$ is
approximated from  below and above by the curves
\be
y_{\pm}(x) =\left[\fr{\pi}{2} +e^{-x}\left(x+\ln \left|x+1+i
\left(\fr{\pi}{2} + e^{-x}(x\pm2\ln(x+1)\right)\right| -c\right)
\right],
\lb{je}
\ee
with $x\in[x_o,\infty]$, $x_o>0$, since
\bea
\lft{Re\,\om (z)|_{y=y_{\pm}(x)} = e^x\cos y_{\pm}(x) +x+\ln\left|x+1+
iy_{\pm}\right|}\sim \nonumber\\
& & \mbox{} \sim c + a_{\pm}\ln\fr{\pi\ln (x+1)}{2(x+1)^2}e^{-x},
\hs{2.5cm}
(x\rightarrow\infty)
\lb{je1}
\eea
where $a_+ =-1$ and $a_- =3$.
The above considertions confirm our expectations that the boundary
of $\partial H_R$ approximates the origin with zero slope, since according
to (\rf{je},\rf{je1}),
\be
\bar{y}\sim \pm \fr{\pi}{2}\bar{r}^2\hs{2.5cm}(\bar{r}\rightarrow\infty),
\ee
where the bared parameters correspond to the mapping
$z\rightarrow\bar{z}=1/z$.

{}From (\rf{oi}) it follows that
\be
\fr{\partial y}{\partial x} =\left(e^x\cos y +1+
\fr{x+1}{(x+1)^2+y^2}\right)/\left(e^x\sin y\right)
\ee
diverges for $y\rightarrow 0$ as expected.
The point $x_o$ at which
$y(x_o)=0$ satisfies approximately the equation
\be
(x_o+1)e^{x_o} = e^c
\lb{ja1}
\ee
and the point $x_m$ at which $y(x)$ takes on its maximum is given
by the equation,
\be
[(x_m+1)^2 + y_m^2]^{1/2} e^{x_m} = e^{c+1}.
\lb{ja2}
\ee
Both (\rf{ja1}) and (\rf{ja2}) are valid asymptotically for
$c$ sufficiently large. Thus, at this limit,
\be
\fr{x_o+1}{[(x_m+1)^2 + y_m^2]^{1/2}}\,e^{x_o -x_m} = e^{c-c-1}=e^{-1},
\ee
i.e., as $x_o\rightarrow \infty (\Leftrightarrow c\rightarrow\infty)$
then $x_m\rightarrow x_o+1$.
Therefore, by the triangle inequality,
\be
|\oint_{\partial H_R} \ldots dz/z|\leq 2\pi \max (\ldots) +
\int_o^R |\ldots| dr/r,
\lb{je3}
\ee
where $\ldots$ stands for any continuous function bounded on
the strip $|Im\, y|\leq \pi$ and $x\in[x_o, \infty)$, $x_o$
being a positive real number.

Transition to more general horns of  the type  $(1,\eta )$  is simply
accomplished by mapping $z\rightarrow z/\eta$. The
theorems could be also modified  for a region of the
asymptotic type $(k,\eta)$ with $k>1$. The situation here
is, however, more involved.
The main result of this subsection is the Nevanlinna-like
theorem for the horn $H_R$ as follows.\vst{0.3cm}\nl
\bth : Let $f(z)$ be analytic in the horn-shaped
          region
          \(H_R :=\{Re\ \omega (1/z)>\omega (1/R)\}\), continuous
          up to the boundary, and satisfy there the estimates
          \[f(z) = \sum _{k=o}^{N-1} a_k z^k + R_N (z)\]
          with
\begin{equation}
          \mid R_N (z)\mid\leq A\mu (N)\mid z\mid^N
\label{h10}
\end{equation}
uniformly in $N$ and $z\in {\bar H}_R$.

Then
\begin{equation}
M(t):=\sum _{n=o}^\infty a_n \fr{t^n}{\mu (n)}
\label{h11}
\end{equation}
          converges  for  $\mid t\mid <1$, and has an analytic
          continuation to the striplike region
          $S_1=\{t\mid dist(t,R_+)<1\}$, satisfying the bound
\begin{equation}
          \mid M(t)\mid\leq K\exp [\exp (\mid t\mid /R)]
\label{h12}
\end{equation}
          uniformly in every $S_{\kappa}$ with $\kappa >1$. The
          analytic continuation
          of $M(t)$ for $t\in (1,\infty )$ is given as follows,
\begin{equation}
M(t) = \fr{1}{2\pi i}\oint _{\partial H_R} F(t/z)f(z)dz/z\ .
\label{h13}
\end{equation}
          Furthermore  f  can  be  represented by the absolutely
          convergent integral
\begin{equation}
          f(x) = \int _o^\infty \exp (-\exp t)M(tx)\ dt\ ,
\label{h14}
\end{equation}
          for any $x\in (0,R)$.\vst{0.3cm}\nl
\lb{hth2}
\eth

The proof of the theorem is rather complicated.
Therefore, it is divided into several lemmas. However, an
importance of $F(t)$ is already seen from the subsequent
argument. If a function $f(z)$ is analytic in the horn $H_R$ and
          continuous in $H_R$, then for any $x\in H_R\cap R$,
\begin{eqnarray}
\lft{f(x) = \fr{1}{2\pi i}\oint _{\partial H_R} f(z)/(z-x) dz}
  \nonumber\\
  && = \fr{1}{2\pi i}\oint _{\partial H_R} f(z)
          dz/z \int _o^\infty  \exp (-\exp t)\, F(tx/z)\, dt
\end{eqnarray}
Note  that  such  a representation of $f(z)$ is impossible by
the Borel  method. Note also that unlike the disc $C_R$ in
the Nevanlinna  theorem  the horn $H_R$ is not a star-like region
anymore. This is in  general the main difference between the regions
          of the asymptotic  type  $(0,\eta )$  and  $(k,\eta)$ with
          $k\geq 1$. This
          difference means that with moment constant
          summability  methods one {\em cannot} recover $f(z)$ from its
          asymptotic
          series in the {\em whole horn} $H_R$ but only  for
          $z\in H_R \cap R$.
Physically this is not, however, a problem, since we are
expanding in real parameters (couplings).

Henceforth we shall follow Sokal's strategy of the proof of the
Nevanlinna  theorem \cite{So}. Lemma \rf{rlm2} provides us with an
          integral  representation  of the monomials
          $t^n /\mu (n)$ for any positive $n$ and $t\geq 1$,
          \begin{equation}
\fr{1}{2\pi i}\oint _{\partial H_R} F(t/z)z^n dz/z = \fr{t^n}{\mu
          (n)},
\label{h19}
          \end{equation}
          where  the  integral  is  taken  counterclockwise along the
          boundary of  $H_R$.  The  formula will be used to express
          $M(t)$
          (an analogue of $B(t)$) in terms of $f(z)$. To   find the
          minimal
          domain  of  analyticity of $M(t)$ we shall need  a  bound
on $F^{(n)}(z_o )$  for $n$ tending to infinity (Lemma \rf{hlm3}).
          Finally,  after the Lemma \rf{hlm4} we shall give all
the lemmas together and complete the proof of Theorem \rf{hth2}.
\vst{0.3cm}\nl

\blm : For any $z_o$ such that $Re\ z_o>0$ and $n>0$,
\begin{eqnarray}
\lft{ |F^{(n)}(z_o)| \leq const\, (n!/\mu (n))}\nonumber\\
& & \exp\left\{\fr{nx_o}{w(n)+1}\left(1+\fr{1}{w(n)}\right)- w(n)+x_o-1
+ {\cal O}(w(n)w'(n))\right\} ,\hs{1cm}
\lb{bac1}
\end{eqnarray}
          where $x_o =Re\ z_o$.
\lb{hlm3}
\elm
{\em Proof} : Firstly,
          \begin{eqnarray}
\lft{F^{(n)}(z_o )/n! = (1/2\pi i)\oint _{C} F(z_o +z)/z^{n+1} dz}
\hs{2cm}\nonumber \\
&& = (1/2\pi i)\oint _C \exp [\omega (z+z_o )-n\ln z] dz,
\label{h20}
          \end{eqnarray}
where $C$ is a simple contour enclosing the origin.
Due to the properties of $\omega (z)$ (see (\rf{om}) and
Lemma \rf{rlm2}) if $| Im\, z_o|> \pi$ and if one takes
$C$ to be a contour of radius $1$  then one can show that
$|F^{(n)}(z_o)|\rightarrow 0$. Hence without any restriction
we can confine ourselves to the region where $| Im\, z_o|\leq \pi$.
In this region we shall evaluate the integral on the r.h.s of
(\rf{h20}) by the saddle point technique.
The saddle  point of the integral is a critical point of the function
\be
h(z) := e^{z_o +z} + z_o+z +\ln(z_o+z+1) -n\ln z,
\ee
i.e., a solution to the equation
%ch (using Lemma \rf{rlm2})
          %
\begin{equation}
 h'(z) = e^{z_o +z} + 1 + 1/(z_o +z+1) -n/z =0 .
\label{h21}
\end{equation}
By direct comparison of (\ref{h21}) with the defining equation
for $w(s)$ (\ref{h17a}), $w(s)e^{w(s)}\\ =s$, one anticipates that
for $n$ sufficiently large the solution $v(n)$ of (\ref{h21})
will be ``very close" to $w(s)$. Therefore we shall
look for the solution $v(n)$ of (\ref{h21}) in the form
\be
v(n)=w(n)-z_o-\delta(n),
\ee
where $\delta(n)$ is an unknown function to be determined. Such
parametrization will be shown to be justified as at the end of our
calculation we shall obtain an asymptotic expansion of $\delta(n)$
for $n\rightarrow\infty$, according to which the
dominant term $-z_o/w(n)\sim -z_o/\ln n\rightarrow 0.$
To prove this we shall slightly rewrite Eq.(\ref{h21}),
\be
v(n)\left(e^{z_o +v(n)} + 1 + \fr{1}{v(n)+z_o +1}\right) = n ,
\ee
and take logarithms of both its sides. Upon the
substitution $v(n)=w(n)- z_o -\delta$
and some manipulations (after one has expanded small terms
in logarithms) one gets,
\be
\ln w -\fr{z_o}{w} -\fr{\delta}{w} +w-\delta+\fr{e^{\delta}}{e^w} +
\fr{e^{\delta}}{e^w (w-z_o-\delta)}\sim \ln n .
\ee
By using the defining relation (\ref{h17a})
for $w(s)$, $\ln w(n) +w(n)=n$, and subsequent multiplication of both
sides of the last Eq. by
$we^{-\delta}/(w+1)$ one obtaines,
\be
\left(\delta+\frac{z_o}{w+1}\right)e^{-\delta} \sim
\left(1+\frac{1}{w}+\frac{z_o+\delta}{w^2}\right)\frac{we^{-w}}{w+1}\cdot
\lb{bu1}
\ee
%ch
By using (\ref{uf4}) one finds that the last fraction on the r.h.s.
of (\ref{bu1}) is nothing but $w(n)w'(n)$. To get $\delta(n)$
in an explicite form one again takes logarithm of both sides
of (\ref{bu1}),
\be
-\delta +\ln\left(\delta+\frac{z_o}{w+1}\right)\sim
\ln(ww') + \frac{w+z_o+\delta}{w^2}\,\cdot
\lb{bu2}
\ee
Dominant terms for $n\rightarrow\infty$ in (\ref{bu2}) are logarithms.
For a moment we shall
parametrize $\delta +z_o/(w+1)=ww' +\epsilon (n) ww'$, where
$\epsilon (n)$ is assumed to be a small number for sufficiently
large $n$. Under this assumption one gets by expanding
small terms in logarithm on the l.h.s. of (\ref{bu2}) that
\be
\epsilon (n)\sim \frac{1}{w(n)} -\frac{z_o}{w(n)+1}\hs{3cm}
(n\rightarrow\infty).
\lb{bu3}
\ee
Thus, the assumption that $\epsilon (n)$ is small for
$n\rightarrow\infty$ is justified and one arrives at the following
expression for $\delta(n)$,
\bea
\lft{\delta(n) = -\frac{z_o}{w+1}+ w(n)w'(n)\left(1+\frac{1}{w(n)}
-\frac{z_o}{w(n)+1}\right) + {\cal O}[(ww')^2]}
\hs{7cm}\nonumber\\
& & \hspace{1cm} (n\rightarrow\infty ).
\lb{bu4}
\eea
This is one of the most important relations to prove Theorem \ref{hth2}.
The asymptotic expansion (\ref{bu4}) justifies the statement made
at the beginning of our calculations that
$\delta(n) ={\cal O}(1/\ln n)\rightarrow 0$ provided $n$ tends to infinity.

Finally, by using (\ref{bu4}), one arrives at the following expression
for the solution $v(n)$ of (\ref{h21}) in the leading order in $n$,
\be
v(n) = w(n) - z_o + \frac{z_o}{w(n)+1} + w(n)w'(n) +
{\cal O}[w'(n)] \hspace{0.5in}(n\rightarrow\infty ).
\lb{uff5}
\ee
Note  that  $z_o +v(n)$  approaches the real positive axis
when $n\rightarrow\infty$ , and a  part of the contour of
the steepest descent nearby the saddle  point  is approximately a
{\em segment of the circle centered at} $z_o$.
After the saddle point evaluation of the integral one finds that
%ch add vv'
\[F^{(n)} (z_o)/n!\sim [v(n)v'(n)/2\pi ]^{1/2}
\exp [\omega (z_o +v(n))-n\ln v(n)].\]

Now we can calculate the behaviour of $n(\ln w(n)-\ln v(n))$
and $e^{z_o+v(n)}-e^{w(n)}$ for $n\rightarrow\infty$.
Both expressions are needed to establish the bound (\ref{bac1}) on
$F^{(n)}(z_o)$.
\[
n(\ln w(n) -\ln v(n)) \ =\ n\ln\frac{w}{v} =
n\ln\left[\frac{e^{z_o+v(n)}+1+1/v(n)}{e^{w(n)}}\right]\ =\hs{2cm}\]
%\nonumber\\
\[ n\ln\left(e^{-\delta}+e^{-w}+\frac{e^{-w}}{w-z_o-\delta}\right)
\sim\ n\left[(e^{-\delta}-1)+e^{-w}+\frac{e^{-w}}{w-z_o-\delta(n)}
\right]\ =\]
%\nonumber\\
\be
n[-\delta(n) +{\cal O}(e^{-w(n)})]\ = n\left[\fr{z_o}{w(n)+1} -
w(n)w'(n)-{\cal O}(w'(n))\right].\hs{1cm}
\lb{bu5}
\ee
\bea
e^{z_o+v(n)}-e^{w(n)} & =& e^{w(n)}(e^{-\delta}-1)
= e^{w(n)}\left[\fr{z_o}{w(n)+1}-w(n)w'(n)-{\cal O}(w')\right] \nonumber\\
&&=\fr{n}{w(n)}\left[\fr{z_o}{w(n)+1}-w(n)w'(n)-{\cal O}(w'(n))\right].
\lb{bu6}
\eea
Thus, when we use that $e^{w(n)}=n/w(n)$,
\[
\fr{F^{(n)} (z_o)\mu (n)}{n!} = (vv'ww')^{1/2}
\]
\[
\exp\left\{e^{v+z_o}+v+z_o+\ln (v+z_o+1)-n\ln v -e^w +n\ln w\right\}=
\hs{0.3cm}\]
%\nonumber\\
\[
\exp\left\{\fr{nz_o}{w(n)+1}\left(1+\fr{1}{w(n)}\right)-w(n)-nw(n)w'(n)
\left(1+\fr{2}{w(n)}\right)+z_o+{\cal O}(w(n)w'(n))\right\}\]
\be
\hs{6cm}(n\rightarrow\infty ).
\label{h22}
\ee
The factor $e^{-w(n)}$ in (\ref{h22}) is produced by $(vv'ww')^{1/2}$
as one can check by using (\rf{uf4}) and (\rf{uff5}).
%On the other hand if we consider the integral
%on the r.h.s. of (\ref{h20}) for $n$ fixed and $z_o$ tending to
%infinity then we have,
%
%\begin{equation}
%\mid F^{(n)} (z_o )/n!\mid\leq\exp [Re\ \omega (z_o +1)].
%\label{h23}
%\end{equation}
%
%Giving  (\ref{h18}), (\ref{h22}) and (\ref{h23})
%together and using the approximate  expression (\ref{h17b}) for $w(n)$ one
%finally gets the statement of the lemma.
$\Diamond$\vst{0.3cm}\nl

\blm :  Let $x\in\,R$ and $z\in\,C$ be independent variables and let
$J_x (z)$ be the integral
\begin{eqnarray}
\lft{J_x (z) := \int_o^1 \exp [-\exp (t/x)]F(tz) dt}\hs{3cm}\nonumber\\
 && = \sum_{n=o}^\infty \frac{(xz)^n}{\mu (n)} \int_o^{1/x}
          \exp (-\exp t)t^n dt .
\lb{oi3}
\end{eqnarray}

          Then we have
\[ \frac{1}{2\pi i}\oint _{\partial H_R} J_x (1/z)z^n dz/z
= \frac{1}{\mu (n)}\int _o^{1} \exp (-\exp t/x)t^n dt ,\]
where  the  integral  is  taken  counterclockwise along the
boundary $\partial H_R$ of $H_R$.
\lb{hlm4}
\elm
{\em Proof} : The essence of the proof is to show that
 the integration (\rf{oi3}) of $F(tz)$  does not spoil
the asymptotic behaviour of $F(z)$ too much.
Note that $x^n\int _o^{1/x}\exp (-\exp t)t^ndt$
behaves like $\exp [-\exp (1/x)]/(nx)$ provided $n$ is sufficiently
large (the asymptotic is given by the end point of integration here).
Thus one anticipates that $x$ will only produce an
overall factor to the asymptotic behaviour of $J_x (z)$ when
$z$ tends to infinity.  This  can be confirmed as in Lemma
\rf{rlm2}.
One uses the Euler-Maclaurin sum formula and finds that for
$\mid Imz\mid\leq\pi$ the asymptotic behaviour is
governed by a saddle point (provided one stays on an appropriate
curve for $\pi/2\leq\mid Imz\mid\leq\pi$ (see the proof of
Lemma \rf{rlm2})), which is here determined by the
          equation
            \[\ln w(s) + 1/s  =  \ln z .\]
          Thus, the contribution $V_s$ of the saddle point is
\[V_s = \frac{(z+1)}{x}\exp\{\exp z[\exp (-\exp (-z)/z)] + 1\}
          \hspace{1.5cm}(z\rightarrow\infty ).\]
By  analogy  with  Lemma \rf{rlm2} one proves that for
$z$ tending to infinity and $\mid Im z\mid>\pi/2 +\varepsilon$,
$\varepsilon\>0$, function $J_x (z)$
tends to zero like $F(z)$ does. The proof of the
          statement of the lemma is then trivial.
It amounts to using the Cauchy integral formula. Note that the
integral  from  $J_x (z)/z$  along  a  segment  of
$\partial H_R^{-1}$, which starts at infinity and terminates  at  some
$z_o$ on the contour, converges even {\em absolutely}.
$\Diamond$\vs{0.3cm}\nl

\subsubsection{Proof of Theorem 5}
\lb{sub2:proof}

% ch the number of Th
{\bf i)}  Under  the  hypotheses of Theorem \rf{hth2} one easily proves
that  the  series (\rf{h11}) converges for
$\mid t\mid <1$. Let us consider the integral $d(t)$,
\begin{equation}
d(t) :=  \fr{1}{2\pi i}\oint_{\partial H_R} F(t/z)f(z) dz/z ,
\label{h24}
\end{equation}
where $t\geq 1$. Two remarks are in order. In contrast to the
Nevanlinna  theorem  one  cannot  use  the  integral on the
r.h.s. of (\ref{h24}) for $0<t<1$, as it is  not possible to satisfy
both conditions that the contour of  integration in (\ref{h19}) be
the contour which tends to  zero  on  the boundary of $tH_r$
for some $r$, and at the same  time  lie in
$H_R$. Whenever $t\not\in R$
the integral  is  {\em identically  zero} (by virtue of Lemma \rf{rlm2}).
Hence, it cannot yield an analytic continuation of $d(t)$ for $t\not\in R$.

{}From  the  properties  of  $F(z)$, it is immediately seen
that the integral on  the  r.h.s. of (\ref{h24}) converges for
$t>1$ absolutely and uniformly  on  any closed subset of
$(1,\infty )$.
For $t=1$  the  integral  converges by virtue of the
Abel-Dirichlet  lemma. Hence, $d(t)$ is a $C^{\infty}$ function on the
interval $(1,\infty )$ and possesses the right derivatives
at the point $t=1$.

To prove that the series (\ref{h11}) converges at $t=1$, we
make use of Lemma \rf{hlm3} and  rewrite $d(1)$ as follows,
\be
d(1) = \sum_{k=o}^{N-1} \fr{a_k}{\mu (k)} +
\fr{1}{2\pi i}\oint_{\partial H_R} F(1/z)R_N (z) dz/z .
\label{h25}
\ee
As  the  integration  on the r.h.s. of (\ref{h25}) runs along the boundary of
$H_R$  on  which $Re\ \omega (1/z)=const\ (=\omega (1/R))$
due to (\rf{je3}) the integral can be bounded from above as follows,
%
%ch
\begin{eqnarray}
\lft{ \left|\fr{1}{2\pi i}\oint_{\partial H_R} F(1/z)R_N
(z)\ dz/z\right|
\leq (A/\pi )e^{\omega (1/R)}\mu (N) \int_{\partial H_R^+}
\exp (N\ln r)\ |dz|/r}\hs{4cm} \nonumber\\
& & \leq 2A\, e^{\omega (1/R)}\,\mu (N)\exp(-N\ln (1/R))
\left(1+\fr{R}{N}\right).\hs{1cm}
\lb{pic1}
\end{eqnarray}
Here $\partial H_R^+$ means that we integrate along $\partial H_R$ in
the first quadrant.
Optimalization of the bound  on the r.h.s. of (\rf{pic}) then amounts
to finding a solution $R=R(N)$ of the equation,
\be
h(R):=-N\,R +\exp(1/R) +1 +\fr{1}{1+1/R} =0.
\lb{joj}
\ee
If one compares (\rf{joj}) with (\rf{h21}) then one finds that
(\rf{joj}) is nothing but (\rf{h21}) with $z_o=0$. Therefore the bound
on the r.h.s. is optimalized by the choice $R=1/v(N)$.
After distorting the contour of integration  up
$\partial H_R$ with $R=1/v(N)$ and using (\rf{bu5},\rf{bu6})
we have,
\[
\left|\fr{1}{2\pi i}\oint _{\partial H_R}
F(1/z)R_N (z)dz/z\right|\leq \hs{7cm}\]
\[
\leq 2A \exp\left\{e^{v} +1 +\ln (v+1)-N\ln v
-e^w +N\ln w-w/2\right\} \hs{2cm}\]
\[
=2A\exp\left\{-Nw(N)w'(N)-w(N)/2-{\cal O}(Nw'(N))\right\}\hs{3cm}\]
\be
\sim 2A\exp (-3w(N)/2)\longrightarrow 0
\hspace{3cm}   (N\rightarrow\infty ),\hs{2cm}
\lb{joj2}
\ee
where we have used that $Nw'(N)\rightarrow 1$ if $N\rightarrow\infty$.
The term $-w(N)/2$ has its origin in the Gaussian prefector
$(2\pi ww')^{1/2}$ of the asymptotic of $\mu(N)$.
Thus,  $M(1)=d(1)$.  The  same  is also true for derivatives, i.e.,
$M^{(n)}(1_-)=d^{(n)}(1_+)$  for any $n\in N$. Indeed, if $f(z)$
satisfies SAC then the same SAC will also satisfy  its
derivatives $f^{(n)}(z)$ (may be in a horn $H_R$ with different $R$). So
\[d^{(n)}(1_+) = \sum _{k=o}^{N-1} c_k /\mu (k) +
\fr{1}{2\pi i}\oint _{\partial H_R} F(1/z)R_N^{(n)}(z)dz/z,\]
where $c_k=(k+n)(k+n-1)...(k+1)a_{k+n}$, and the integral can be
estimated in the same manner as above.

To  determine the minimal region of analyticity we express $d^{(n)}(t)$ as
follows:
 \[d^{(n)}(t) = n!a_n /\mu (n) +
 \fr{1}{2\pi i}\oint _{\partial H_R}F^{(n)}(t/z)R_{n+1}(z)z^{-n}
 dz/z.\]
Now, using Lemma \rf{hlm3}, one finds that
\begin{eqnarray}
\lft{\mid d^{(n)}(t)\mid\leq \mb{const} F(t/R)\ n!}\hs{1cm}\nonumber\\
&&
 \exp [(t/R)(n/\ln n+n/\ln ^2 n)+2\ln w(n)+ {\cal O}(1/w(n))].
\label{h26}
          \end{eqnarray}
One may justify that
          \[\sum _{n=o}^\infty d^{(n)}(t_1 )(t-t_1 )^n /n! =
          \sum _{n=o}^\infty d^{(n)}(t_2 )(t_2 -t)^n /n!,\]
whenever $1<t_1\leq t\leq t_2$, and $\mid t-t_i\mid <1$, where $i=1,2$.

Thus $M(t)$,
\[ M(t) = \left \{\begin{array}{ll}
  \sum _{n=o}^\infty a_n t^n/\mu (n) & -1<t\leq1,\\
  d(t)\hs{0.8cm}                    & t\geq 1 ;
\end{array}
\right. \]
is  shown  to  be analytic at least in the striplike region
$S_1$ (see Fig. \rf{fg6}).
\bfg
\vs{8cm}
\caption{{\em Minimal region of analyticity of $M(t)$.}}
\lb{fg6}
\efg
Bound  (\ref{h12}) on $M(t)$ follows immediately from the
relation (\ref{h26}).

{\bf ii)}  To  prove the relation (\ref{h14}), let us consider the
          integral
          \begin{equation}
\fr{1}{x} \int _o^\infty\exp [-\exp (t/x)]M(t)dt ,
\label{h27}
          \end{equation}
          for any $x\in (0,R)$. The integral,
\[\int _1^\infty \exp [-\exp (t/x)]\ \fr{1}{2\pi i}
          \oint _{\partial H_R} f(z)
          F(t/z)\ dz/z\ dt,\]
is  absolutely  convergent.  Therefore, the r.h.s. of (\ref{h14})
(the integral (\ref{h27})) can be recast into the form,
\begin{eqnarray}
\lft{f(x) - \fr{1}{2\pi ix}\oint _{\partial H_R}
J_x (1/z) f(z)\ dz/z}\hs{4cm} \nonumber\\
&& + \fr{1}{x}\sum _{n=o}^\infty \fr{a_n}{\mu (n)}\int _o^1\exp
          [-\exp (t/x)]\ t^n\ dt.
\end{eqnarray}
Now we shall show that the last two terms cancel
          each other. Indeed,
using  Lemma \rf{hlm4}, one finds
\begin{eqnarray}
\lft{\fr{1}{2\pi i}\oint _{\partial H_R} J_x (1/z)f(z)dz/z =
\sum _{n=o}^{N-1} \fr{a_n}{\mu (n)}\int _o^1\exp [-\exp
(t/x)]\ t^n\ dt}\hs{4cm} \nonumber\\
&&
+ \fr{1}{2\pi i}\oint _{\partial H_R} J_x (1/z)R_N (z)\ dz/z\ .\nonumber
\end{eqnarray}
To show that the last integral vanishes in the limit
$N\rightarrow\infty$ one optimalizes  the  bound  on  the integral.
To do this one
chooses similarly like in, (\rf{pic}-\rf{joj2}),
the contour of integration to be
$\partial H_{1/v(N)}$. Then, in virtue of (\ref{h17a},\rf{h17b},\rf{r18}),
%ch the references
%
\beas
\left|\fr{1}{2\pi i}\oint _{\partial H_R} J_x (1/z)R_N
          (z)dz/z\right|
\leq\frac{2A}{x}e^{-3w(N)/2} \longrightarrow 0
                 \hspace{1cm}(N\rightarrow 0).
\eeas
$\Diamond$\vst{0.3cm}\nl

 \bcl : Let

          1)  $M(t) := \sum _{n=o}^\infty a_n t^n /\mu (n)$ has
          nonzero radius of convergence
          and can be analytically extended on the real positive axis;

          2) \(\mid M(t)\mid\leq const\ \exp (\exp (t/R))\) on the
          axis.

          Then
 \[f(x) := \fr{1}{x}\int _o^\infty\exp (-\exp (t/x))M(t) dt \]
          converges absolutely for $x\in (0,R)$ and
  \[f(x)\sim\sum _{n=o}^\infty a_n x^n
          \hspace{1cm} (x\rightarrow 0_+).\]
\lb{hcl1}
\ecl
{\em Proof} : Use the Laplace method to evaluate the integral.
$\diamondsuit$\np

\subsection{Comparison of the new summability method with the
Borel method}
\lb{sub:comp}

Using a  recent
generalization of the Borel summability method \cite{M} and the
Nevanlinna theorem \cite{M1} one can easily find
series
which violate SAC in a horn-shaped region $H_R$ but they are Borel
summable (see
{\em Example 1} below). On the other
hand one can find series which are formally Borel nonsummable but
satisfy
SAC in a suitable kidney-shaped region (see {\em Example 2}). This
exemplifies the statement in the Introduction
that not the Borel (non)summability but the validity (violation)
of SAC prevents (indicates) the presence
of nonperturbative effects which may in some cases destabilize a
ground state. The first shows a  series
which violates SAC if summed up to a function analytic
in the horn $H_R$ \cite{M1}, while it satisfies them if
summed up to a function
regular in a disc tangent to imaginary axis at the origin.
The main method for constructing them is to
use the new AMCSM based on the momenta (\ref{m3}) as a method for
{\em localization of singularities in the
complex plane}. Indeed, this method has a nice and rather feasible
advantage as compared with the Borel one - {\em it does not only see
beyond singularities}. The problem of finding singularities of a
function $f(z)$ which is defined by a convergent power series is then
reduced to the
problem of finding the asymptotic behaviour of its moment transform.
To find the asymptotic behaviour of the moment transform the
{\em Euler-Maclaurin formula} will be frequently used. It reduces the
above problem
to the problem of finding the asymptotic behaviour of an integral.
By our general results \cite{M,M1} we know that in general the $\mu$-transform
may have dangerous asymptotic behaviour (i.e., which may prevent
summability of its Taylor expansion) at worst on several (possibly infinite)
radial rays. This should be contrasted with the Borel
method - any singularity of $f(z)$ results in that the Borel transform
of its Taylor series grows up very strongly (such that the Borel
integral diverges) not only on the radial ray which passes through the
singularity but also in the whole sector bisected by this ray and with
the opening angle $\pi$.\vst{0.3cm}\nl

\begin{E} Let $S_1$ be a power series \(\sum _o^\infty a_ng^n \) with
\( a_n = (-1)^n \mu ^2 (n)\),
Then $S_1$ is Borel summable and satisfies SAC in the disc $K(0,R)$
 but is not
$\mu $-summable and violates SAC in the horn $H_R$.
\end{E}
{\em Proof}: The Borel transform of $S_1$ is an entire function and hence
it is obviously  analytic on a  strip
          $S_{\rho }:=\{t\mid dist(t,R_+ )<1/\rho \}$ for any
 $\rho $, which is the first hypothesis of the Nevanlinna
theorem \cite{N,M,M1}. Now we use the
 Euler-Maclaurin  sum formula \cite{J},
\[ B(t) = \int_{\sigma}^\infty e^{s(\ln t+i\pi )}\frac{\mu ^2(s)}
{\Gamma(s+1)} ds +
O(\mid t^\sigma\mid ) \hspace{1in}(t\rightarrow\infty),\]
\noindent where $\sigma $ is a real number, $-1<\sigma <0$.
One finds that a saddle point of the Euler-Maclaurin
integral is at $s/w^2(s)=-t$. Thus the asymptotic of the integral is
dominated by a contribution of the end point and we arrive at the result
\[ B(t) \sim  O(\mid t\mid ^{\sigma}) \hspace{1.5in}(t\rightarrow
\infty ), \]
in any sector with nonzero opening angle
$\mid\arg t\mid<\pi/2$.
Therefore the hypotheses of the Nevanlinna theorem are satisfied, $S_1$ is
Borel summable, and its Borel sum satisfies SAC in the disc $K(0,R)$.

On the other hand, the generalized $\mu $-transform $M(t)$,
\[ M(t) := \sum _o^\infty \frac{a_n t^n}{\mu (n)} =
\sum _o^\infty (-1)^n t^n \mu (n), \]
\noindent is undefined since it has zero radius of convergence,
i.e., the summability method for the
horn-shaped region $H_R $ in this case fails.
$\diamondsuit$\vst{0.3cm}\nl

\begin{E} Let us consider the series $S_2 = \sum _o^\infty a_n g^n $
 with
\[ a_n = (-1)^n \Gamma (\alpha n +1), \]
\noindent where $\alpha = 1+\varepsilon $, $1>\varepsilon >0$, and
$\Gamma$ is the usual gamma function.

Then $S_2 $ cannot be summed up to a function maximal region the
analyticity of which is a disc tangent to the imaginary axis at the origin.
Hence it is not Borel summable and violates SAC in this region.
Nevertheless, it can be summed up to a function whose minimal region
of analyticity contains a kidney shaped region $K(\lambda ,R)$,
\[ K(\lambda ,R) = K^\lambda (0,R) := \{z\mid z^{1/\lambda}\in K(0,R)\},
\]
having an opening angle $\Theta = \lambda\pi$ with $\lambda\geq\alpha$,
$K(0,R)$ being the disc $Re(1/z)>1/R$, i.e., $S_2$  is
summable and satisfies SAC in
this region.
\end{E}
For the proof see \cite{GGS,N}. $\diamondsuit$\vs{0.3cm}\nl

Eventually, we shall give an example just opposite to {\em Example 1}.
\begin{E} Let $S_3$ be a power series with  co\-ef\-fi\-cients
\( a_n = a^n\mu (n) \), where \mbox{\(a=\mid a\mid e^{i\theta}\)} is a
complex number
with \(0<\theta <\pi /2\). Then the series $S_3$  is $\mu $-summable
but not Borel summable.
\end{E}
{\em Proof} : The Borel transform of $S_3$ is an entire function.
The position of a saddle point of its Euler-Maclaurin integral
representation is given by equation
\[ w=\ln (aue^{i(\phi +\theta)}),\]
where $t=ue^{i\phi }$, which means that the contribution $V_s$ of the
saddle point is
 \[ V_s \approx \exp[au\exp(i(\phi +\theta))(\ln(au)-1+i(\phi +\theta))]
 \hspace{1in}(u\rightarrow\infty).\]
Hence, the Borel integral diverges whenever $\theta<\pi/2$.
On the other hand
$\mu $-transform $M(t)$ can be calculated exactly,
\[ M(t) = \sum_{n=o}^\infty\frac{a_n}{\mu (n)}t^n = \frac{1}{1-at} , \]
and $S_3$ is transparently $\mu $-summable (see \cite{M,M1}).

If $3\pi/2\geq\theta\geq\pi/2$ then one can show that
\[\int _o^\infty e^{-t}\sum_{n=o}^\infty\frac{a_n}{n!} (tz)^n dt =
\int_o^\infty e^{-e^t}\sum_{n=o}^\infty\frac{a_n}{\mu (n)}(tz)^n dt,\]
since the real positive axis belongs to the Borel polygon of $M(z)$.
$\diamondsuit$\vst{0.3cm}\nl

The above examples demonstrate that there exist series which being Borel
summable are not $\mu$-summable and vice versa. They also show  that
$\mu $-method \cite{M,M1}, in combination with the
Euler-Maclaurin sum formula, is a powerfull tool for looking for
singularities of  analytic continuation of convergent power series in
the complex plane, since
in contrast to the Borel method  $\mu $-method {\em does not only see beyond
singularities}.
%The method can be implemented numerically
%as well, since the
\np

\subsection{Derivation of strong asymptotic conditions for variety
of regions}
\lb{sub:dar}

To generalize SAC let us start with some
definitions.
\bdf Let S be the halfplane $Re\, z > 0$,  and  $C_R$ a disc
centered at the origin with
radius R. Denote $\bar{S}:=S\backslash C_R$ which will be called
the base region. A  domain D of the complex
plane C will be called
of asymptotic  type $(k,\eta )$, $k\geq 1,$ if there exist $R>e_{k-1}(1)$
such  that $D=1/[\eta \ln_k(\bar{S})]$,
where $e_k $ is
the k-fold exponential, $e_o (x):=x,\  e_k (x)~:=\exp(e_{k-1}(x)),$
and $\ln_k$ is the k-fold  logarithm,
$\ln_1 (z):=\ln z,\  \ln_k (z)~:=\ln (\ln_{k-1}(z))$.  A
          domain  D of the complex plane will be called of asymptotic
          type  $(0,\eta )$ if $D=1/\bar{S} ^{\eta}$, where
$\bar{S}^{\eta}:=\{z\mid z^{1/\eta} \in\bar{S}\}$.
\vst{0.3cm}$\diamondsuit$\nl
\edf
To establish SAC for these
regions we shall start with the regions of asymptotic type
$(1,\eta  )$.
The main result of this subsection is the following theorem.
\vst{0.3cm}\nl
\bth  Let f(z)

 1)  be  regular in a region D of the asymptotic type
          $(1,\eta  )$
          and continuous in its closure;\nl

          2) have in D an asymptotic expansion
          \begin{equation}
          f(z) =  \sum_{n=o}^{N-1} a_n z^n + R_N (z) .
\lb{eq:bumas}
\end{equation}

If $R_N (z)$ satisfies the bound
\begin{equation}
\mid R_N (z) \mid  \leq A\rho ^N \mid z\mid ^N
\exp[-\exp w(N)+N\ln w(N)]
\label{eq:b}
\end{equation}
with $\rho\leq\eta $, uniformly in
$z\in\bar{D}$  and  $N$, $w(s)$ being the solution of
$w(s)\exp[w(s)]\\ =s$ \cite{M,M1}, then the asymptotic series
(\ref{eq:bumas})
determines $f(z)$ uniquely. The condition (\ref{eq:b}) is strong
in the following sense: if it is known that (\ref{eq:b}) is only
fulfilled with some $\rho >\eta$, uniformly in $z\in\bar{D}$ and N,
then there exists a nonzero function with the trivial asymptotic
expansion and the asymptotic series (\ref{eq:bumas}) does not determine
f(z) uniquely.$\Diamond$\vst{0.3cm}\nl
\lb{sacth}
\eth
Note that $\exp\{-\exp w(N)+N\ln w(N)\}$ is, up to the factor
$[2\pi w(N)w'(N)]^{1/2}$ which arises from the Gaussian integration
around a
saddle point, precisely the asymptotic for large $N$ of
the moment function $\mu (N)$ (see (\ref{h17a}) above).
For  regions of, roughly speaking, asymptotic type  $(1,\eta )$ SAC have
been established in \cite{M1}  in connection with a summability method
for the horn-shaped region $H_R$. Unfortunately, generalization of this
proof to more sharper horns seems to be cumbersome.
A rather brief derivation of SAC based on a slight modification of the
Phragmen-Lindel\"{o}f theorem and which can be straightforward
generalized to
sharper horn-shaped regions has been given in \cite{M2} and is briefly
repeated in  Appendix.
One has only to replace $w(s)$ by
          $w_k (s)$,
          $w_k (s)$ now being the solution of
          \[ w(s)\exp\{w(s)+...+e_{k-2}[w(s)]+e_{k-1}[w(s)]\} = s \]
(see \cite{M}). One can also consider other types of asymptotic regions
which
may be obtained, e.g., by more involved combinations of scaling
$z\rightarrow z/\eta$ and conformal
mappings like $z\rightarrow z^{1/\alpha}$ of $\bar{S}$, as well as some
other base
regions, etc. In general, however, to any given asymptotic region such a
moment sequence $\tilde{\mu }_k(n)$ will
correspond that fulfilling a bound
like (\ref{eq:b}) will ensure SAC in the region. We used the above
definition of the asymptotic types since the maximal region of
analyticity of four dimensional
renormalizable massless field  theories  has  been suggested
to be  at best just a region  of  the asymptotic
type $(1,\gamma )$, where $\gamma $ depends on the first
two coefficients
$\beta _1 , \beta _2 $ of the $\beta $-function and on the definition of the
coupling\footnote{For  't  Hooft's
coupling, $\gamma$ is just $\mid\beta_1\mid$.}
\cite{'tH1,Wi,KR,GK}. This means that probably  there is {\em no reason
to look at the Borel  transform of the four dimensional
renormalizable field  theories} since  {\em the convergence} of the Borel
integral {\em contradicts} the horn-shaped region of analyticity
\cite{M,M1,W}
any way that the Borel transform on the real positive  axis  is  defined
(e.g., in a  {\em distributional  sense} \cite{GGM}). In connection with
Theorem \rf{sacth} this means
that these theories (without UV cutoff) {\em cannot} satisfy SAC since their
analyticity in the complex coupling constant plane {\em is not} compatible with
the divergence of order $(n!)^\varepsilon$  of their perturbation theory
no matter how small  $\varepsilon >0$ is.

Now we want to show
an  important property of the class ${\cal K}$ of  function
which obey the hypotheses of Theorem \rf{hth2} and \rf{sacth}.
Using asymptotic behaviour of $w_k(s)$ \cite{M} and the method of
\cite{M1}
one finds that the SAC we
are  given  preserve nonlinear perturbation conditions such
as unitarity  of the Feynman series. Indeed, the following Lemma
holds.\vst{0.3cm}\nl

\blm : The class of function ${\cal K}$ is closed under
          product, i.e., if $f_1 (z)$ and $f_2 (z)$ are two  functions
          from {\cal K}, then
          \(g(z):=f_1 (z).f_2 (z)\in {\cal K}\) .
\lb{hlm5}
\elm
{\em Proof} : To prove the lemma it is sufficient to prove that
          \begin{equation}
          \lim _{N\rightarrow\infty}\sum _{k=o}^N \mu (N-k)\mu
          (k)/\mu (N) <\infty .
\label{h28}
          \end{equation}
{}From  the  Lemma \rf{rlm1} one can derive that if $N$ is sufficiently
          large then for $q<N/2$,
          \[\mu (N-q)/\mu (N)\sim\exp [-q\ln\ln N
          -q/N-(q+q^2 )/(N\ln N)].\]
Thus, provided one choses a fixed $j$ such that the asymptotic formula to
estimate $\mu (s)$ for $s\geq j$ can be used,
\begin{eqnarray}
\sum _{k=o}^N \mu (N-k)\mu (k)/\mu (N)\leq
   2\sum _{k=o}^{[\frac{N}{2} +1]}\mu (N-k)\mu (k)/\mu (N)\nonumber\\
\leq  const\sum _{k\geq j}^{[\frac{N}{2} +1]}
          \exp\left\{ -\ln\ln N \left[\fr{\ln k}{\ln\ln N} +
          k \left(1-\fr{\ln\ln k}{\ln\ln N}\right)\right]\right\}.
\nonumber
\end{eqnarray}
The last bound gives transparently convergent  series  in the limit
$N\rightarrow\infty$,    since    its    last    term
behaves like \(1/[N\exp (N\ln 2/\ln N)]\) . $\Diamond$\vs{0.3cm}\nl

The next two lemmas deal with the important case of
perturbation series with equal sign and alternating sign of
coefficients \ct{BGZ,GP,MO,GM,BGZ1}.
To formulate the lemmas let us firstly consider a general analyticity
region $D$ to which corresponds momenta
$\tilde{\mu}_k( n)$. Then one can meet the subsequent situations:
\vst{0.3cm}

\(\begin{array}{ll}
a) \forall t>0 \ :\  a_nt^n/\tilde{\mu} _k ( n)\rightarrow\infty &
\hspace{1in}
(n\rightarrow\infty );\\
b) \forall t>0\ :\ a_nt^n/\tilde{\mu}_k( n)\rightarrow 0 & \hspace{1in}
(n\rightarrow\infty);\\
c) \exists t\neq 0\ :\ a_n t^n/\tilde{\mu}_k( n)\rightarrow K\neq 0
& \hspace{1in}(n\rightarrow\infty).
\end{array}\)\vst{0.3cm}\nl
\blm Let S be a divergent power series with equal sign
coefficients. Then S violates SAC in any analyticity region D.
\lb{slm2}
\elm
{\em Proof} : In case a) the moment constant transform $\tilde{M}_k (t),$
\begin{equation}
\tilde{M}_k (t):=\sum _{n=o}^\infty a_nt^n/\tilde{\mu} _k(n),
\label{eq:trans}
\end{equation}
does not exist. In case b)
the moment constant transform is an entire function and hence defined
for all
complex $t$. It is clear that the maximum modulus of $\tilde{M}_k (t)$ for
$\mid t\mid\leq x$ is just $\tilde{M}_k (x)$. From simple
relation between the Taylor series coefficients and the maximum modulus growth
of entire functions based on the Cauchy integral formula \cite{J}
one can prove that the generalized moment constant sum does not exist,
since the integral
\begin{equation}
\int _o^\infty \exp(-e_k(t))\tilde{M}_k (zt^\rho )dt   \label{eq:sum}
\end{equation}
diverges for all real $z>0$. The same is also true  in case c) since
$\tilde{M}_k (t)$ is singular on the real positive axis. $\Diamond$
\vst{0.3cm}\nl

\blm Let S be a divergent power series with alternating sign
regular coefficients $a_n$, i.e., there exist an analytic function a(s)
in the
complex halfplane $Re\, s>\sigma$  such that $a_n=(-1)^na(n)$. Let $a(s)$
have parametrization  $a(s)=\exp (s\ln b(s))$ with
\( sb'(s)/b(s)\sim O(1)\) (or o(1)) when $s\rightarrow\infty$. Let D be the
analyticity domain to which
the momenta $\tilde{\mu }_k ( n)$ correspond. Then in  cases b) and c)
S satisfies SAC in D.
\lb{slm3}
\elm
{\em Proof} : In case b) the moment constant transform $\tilde{M}_k (z)$
is an entire function so the requirement of analyticity of $\tilde{M}_k
 (z)$ to sum S is satisfied.
The Stieltjes moment function $\tilde{\mu}_k(s)$ can be shown
to be
nonzero in the complex halfplane $Re\, s>\sigma'$ for $\sigma'$
sufficiently
large \cite{M}. So, we can again use the Euler-Maclaurin sum formula to
find asymptotic behaviour of $\tilde{M}_k (z)$ for $z\rightarrow\infty$.
A contribution of the saddle point is proportional to
\[\exp \{e_k[\tilde{w}_k(s(z))]\}\approx \exp [e_k(-z)]\sim O(1)
\hspace{1in}
(z\rightarrow\infty),\]
\noindent where $s(z)$ is determined by the saddle
point equation
\(\tilde{w}_k(s)\exp [-a'(s)/a(s)]=-z:=-re^{i\phi}\). However, the saddle
point does not lie on the principal sheet and the contour of
integration cannot probably  be deformed in such a way that the
asymptotic behaviour
of the Euler-Maclaurin integral be governed by it. Instead of that the
contour
of integration can be deformed in such  a way that the asymptotic
behaviour
of the integral can be shown to be governed by the end point of
integration
\cite{J}. In any case, however,
\begin{equation}
\tilde{M}_k (z) \sim O(\mid z\mid ^{\tilde{\sigma}} ) \hspace{1in}
(z\rightarrow\infty),
\end{equation}
where $\tilde{\sigma}=\max \{\sigma, \sigma'\}$, i.e., the generalized
moment constant sum (\ref{eq:sum}) exists. In case
c) $\tilde{M}_k (z)$ is a meromorphic function regular on the real
positive
axis. To see this one may apply $\mu $-method \cite{M,M1} on the series
(\ref{eq:trans}). The $\mu $-transform of $\tilde{M}_k (z)$
is an entire function with alternating sign
regular coefficients and this case can be treated as  case b).
Using the Euler-Maclaurin integral representation of the $\mu$-transform
$\tilde{M}_k (z)$ one finds
that $\mu $-sum converges $\forall\, t\geq 0$ even in a sector and is
absolutely
bounded therein. Therefore $\tilde{\mu}_k$-sum also exists in this sector
and satisfies SAC in $D$. $\Diamond$\np

\section{Applications}
\lb{sec:appl}
\subsection{Summability methods and the Rayleigh-Schr\"{o}dinger
          perturbation theory}
\lb{sub:sum}
\subsubsection{Preliminaries}
\lb{sub2:pre}

As  some  example of application of the summability methods
we  shall consider  the  Rayleigh-Schr\"{o}dinger
perturbation theory \cite{RS,K}. We shall restrict ourselves to
the case of {\em relatively bounded perturbations} \ct{K}.
Let $T$ and $A$ be operators with the same domain space {\bf X} (but
not necessarily with the same range space) such that
${\bf D}(T)\subset {\bf D}(A)$ and
\be
\| Au\| \leq a\|u\| +b\| Tu\|,\hs{1cm} u\in{\bf D}(T),
\lb{relb}
\ee
where $a$ and $b$ are nonnegative constants. Then we shall say that $A$ is
{\em relatively bounded with respect to} $T$ or simply $T-bounded$. The
smallest lower bound $b_o$ of all possible constants $b$ in (\rf{relb})
will be called the {\em relative bound} of $A$ with respect to $T$ or simply
the $T-bound$ of $A$. If $b$ is chosen very close to $b_o$, the other
constant $a$ will in general have to be chosen very large; thus it is in
general impossible to set $b=b_o$ in (\rf{relb}).
Obviously a bounded operator $A$ is $T$-bounded for any $T$ with
${\bf D}(T)\subset {\bf D}(A)$, with $T$-bound equal to zero.

To consider relatively bounded perturbations is very natural since
{\em closedness, bounded invertibility, selfadjointness,
as well as some other properties}  {\em are stabile
under relatively bounded
perturbation} \ct{K}.\vst{0.3cm}\nl
\blm
Let T and A be operators from {\bf X} to {\bf Y}, and let A be
T-bounded with T-bound smaller than 1. Then S:=T+A is closable
if and only if T is closable; in this case the closures of T and S
have the same domain. In particular S is closed if and only if T is.
\lb{ulm}
\elm
Proof is rather simple. We shall use the treatment exposed, e.g.,
in \ct{K}.
In the inequality (\rf{relb}) we may assume that $b<1$. Hence
\be
-a\| u\| +(1-b)\| Tu\|\leq\| Su\|\leq a\| u\| +(1+b)\| Tu\|,
\lb{relb2}
\ee
for ${\bf D}(T)$.
Let us recall that $T$ is closable if and only if
$u_n\in{\bf D}(T)$, $u_n\rightarrow 0$ and $Tu_n\rightarrow v$ imply
$v=0$. Applying the second equality of (\rf{relb2}) to $u$ replaced
by $u_n-u_m$, we see that a $T$ convergent sequence $\{u_n\}$ (that is
a convergent sequence $\{u_n\}$ for which $Tu_n$ is also convergent)
is also $S$-convergent. Similarly from the first inequality an $S$-convergent
sequence $\{u_n\}$ is $T$-convergent. If $\{u_n\}$ is $S$-convergent
to 0, it is $T$-convergent to 0 so that $Tu_n\rightarrow 0$ if $T$ is
closable; then it follows from the second inequality of (\rf{relb2}) that
$Su_n\rightarrow 0$, which shows that $S$ is closable. Similarly, $T$ is
closable if $S$ is. $\Diamond$\vst{0.3cm}\nl

Let $T$ and $A$ be operators from ${\bf X}$ to ${\bf Y}$,
$A$ being $T$ bounded (\rf{relb}).
If $T^{-1}$ exists and is a bounded operator from ${\bf Y}$ to
${\bf X}$ then $AT^{-1}$ is an operator on ${\bf Y}$ to ${\bf Y}$
and is bounded by
$$
\| AT^{-1} v\|\leq a\|T^{-1}v\| + b\|v\|\leq (a\|T^{-1}\| +b)\|v\|.
$$
If
\be
a\|T^{-1}\| +b< 1
\lb{a1}
\ee
one can also show {\em stability of bounded invertibility} under
relatively bounded perturbation.
Indeed if (\rf{a1}) is valid then $S=T+A$ is automatically closed by
Lemma \rf{ulm}. One has
$$
S = T+A =(1+AT^{-1})T.
$$
and thus
$$
\| S^{-1}\|\leq\fr{\|T^{-1}\|}{1-a\|T^{-1}\|-b\|}\cdot
$$
In the Hilbert space one can show that selfadjointness is also stabile
under
relatively bounded perturbations. If $T$ is selfadjoint and $A$ is
symmetric and $T$ bounded with $T$-bound smaller than $1$, then $T+A$ is
also selfadjoint. In particular $T+A$ is selfadjoint if $A$ is bounded
and symmetric with ${\bf D}(A)\supset{\bf D}(T)$ \ct{K}. Note that the
assumption that the bound be smaller than $1$ cannot be dropped in
general. If $T$ is unbounded and selfadjoint, and $A=-T$, then
$T+A$ is a proper restriction of the operator $0$ and is not selfadjoint
\ct{K}.

An important example of a relatively bounded perturbation provides
the Schr\"{o}dinger operator in the $3$-dimensional euclidean space $R^3$
for a system of $s$ particles interacting with each other by the Coulomb
forces. In this case the formal Schr\"{o}dinger operator is the
$3s$-dimensional Laplacian $-\triangle$ and the perturbation $V(x)$ has
the form
\be
V(x)=\sum_{j=1}^s\fr{e_j}{r_j} + \sum_{j<k}\fr{e_{jk}}{r_{jk}},
\ee
where $e_j$ and $e_{jk}$ are constants and
\beas
r_j &=& (x_{3j-2}^2 +x_{3j-1}^2 +x_{3j}^2)^{1/2},\\
r_{jk}&=& [(x_{3j-2} -x_{3k-2})^2 +(x_{3j-1}-x_{3k-1})^2
+(x_{3j}-x_{3k})^2]^{1/2}.
\eeas
It can be proved that the minimal operator $\dot{T}$ constructed from the
formal operator $-\triangle$ is essentially self-adjoint with the
self-adjoint closure $H_o$. If $V$ denotes the maximal multiplication
operator $V(x)$ then it can be shown that $V$ is relatively bounded
with respect to $H_o$ as well as to $\dot{T}$ {\em with relative bound equal
to zero} \ct{K}.

\subsubsection{Setting up the problem}
\lb{sub2:spe}

Henceforth we restrict ourselves to the ensuing problem.
Let  $H_o$ be a closed linear operator acting in a Banach
space and let $V$ be another closed linear operator acting in
this space,  which is relatively bounded with respect to
$H_o$. In addition let the bound satisfy (\rf{a1}) with $T$ replaced
by $H_o$.
We  shall  be interested in the resolvent operator
          $R_\lambda (z):= (H_o  +  \lambda V  - z)^{-1}$ defined
          for $z$ not in the spectrum of
          $H_{\lambda} :=  H_o + \lambda V$ ($\lambda\in C$).
          Usual perturbation theory starts from the identity
          \begin{equation}
          R_{\lambda}(z) = R_o (z)[1 + \lambda VR_o(z)]^{-1}
 \lb{m4}
          \end{equation}
          by developing the geometric series
          \begin{equation}
          [1 + \lambda VR_o (z)]^{-1} = \sum _{n=o}^{\infty}
          (-\lambda )^n [VR_o (z)]^n .
\lb{m5}
          \end{equation}
Under our  assumptions, $VR_o (z)$ is a bounded operator, so
this series will converge in norm for $\|\ld VR_o (z)\| <1$.
If $V$ is bounded then the condition can be replaced by
$\|\ld V\|< d(z)$, where $d(z)$ is the distance of $z$ from the
spectrum of $H_o$.

Let  us  suppose  that  there  exists  a finite system
          $\Sigma '(H_o )$ of eigenvalues of $H_o$, separated from
          the rest of the
          spectrum of $H_o$ by a closed curve $\Theta$ encircling
          $\Sigma '(H_o)$. Then
          there is a convex region $\Lambda$ of the  complex  plane,
          containing the origin,
          such  that for all $\lambda\in\Lambda$ the  spectrum
          $\Sigma (H_{\lambda})$  of  $H_{\lambda}=H_o+\lambda V$  is
          likewise separated by $\Theta$ into a
          part $\Sigma '(H_{\lambda})$ and a remainder. The
          eigenvalues in $\Sigma '(H_{\lambda})$ are
          analytic in $\lambda$ with only  algebraic singularities
\ct{K}. The usual Rayleigh-Schr\"{o}dinger  perturbation theory is
obtained by substituting (\ref{m4}-\ref{m5}) in the expression for
$P_{\lambda}$,
\be
P_{\lambda} = \frac{1}{2\pi i}\oint_{\Gamma} R_{\lambda}(z)dz,
\lb{prt}
\ee
the  {\em projection  operator} onto the subspace associated with
the eigenvalues $\varepsilon _{\lambda}$ branching off from
$\varepsilon _o$ in $\Sigma '(H_o),$ where
$\Gamma$ is a contour  encircling  $\varepsilon _{\lambda}$
but no other points of $\Sigma '(H_\lambda )$ (see Fig. \rf{fg7}).
\bfg
\vs{8cm}
\caption{{\em Perturbation of eigenvalues.}}
\lb{fg7}
\efg
In the physical language if the contour $\Gamma$ encloses all
energy levels below the Fermi surface then $P_\ld$ is nothing but the
Fermi projector or {\em density matrix} at zero temperature,
respectively.

To extend  the region of convergence some summability method can be used
\cite{RS,44}. Reeken used the Borel summability
with $\beta =1$. As  we  have shown, $\mu$-method has a
bigger region of convergence, so that
an application of the method will provide a futher extension of
the Rayleigh-Schr\"{o}dinger
perturbation theory.  Let $A$ be a bounded operator acting in
a Banach space. Consider an {\em operator-valued function} $f(z)$,
$$f(z) = \frac{1}{(1 + zA)}\cdot $$
Because of the Theorem \rf{rth1} (see also \cite{M}), it is true that
\be
f(z) = \frac{1}{(1+ zA)} = \int _o^{\infty} \exp (-\exp t)
 \sum _{n=o}^{\infty}\ \frac{(-zAt)^n}{\mu (n)}\ dt
\lb{pac}
\ee
for $z\in MLS(f)$.  The r.h.s. of (\rf{pac}) can be defined by means of
the uniform
convergence of Theorem \rf{rth1} as a limit of entire functions. Due to
Theorem \rf{rth1} for any  compact set $K\subset
MLS(f)$ and each $\varepsilon$ there is $t_o>0$ and an integer $N$ such that
$$
\left|f(z)-\sum_{n=o}^N \fr{z^n}{\mu (n)}(a_n \int_o^{t_o}
e^{-e^t}t^n\, dt)\right|
<\varepsilon
$$
uniformly in $z\in\,K$.
Thus
$$
\int_o^\infty e^{-e^t} \sum_{n=o}^\infty \fr{(-ztA)^n}{\mu(n)}\, dt=
\lim_{t_o\rightarrow\infty}\lim_{N\rightarrow\infty}
\sum_{n=o}^N (-zA)^n \int_o^{t_o} e^{-e^t}t^n\, dt
$$
as a {\em limit of entire functions}.
Any operator valued function can be defined on its $MLS$ in this
manner. Thus this
provides an alternative to the Dunford-Schwartz integral.
To define an operator valued
holomorphic function $f(z)$ on $MLS(f)$ as a limit of entire
functions one can also use some other AMCSM \ct{Mo,H}.

The  singularities of $f(z)$ are just
such $z$ that $-1/z$ is  from  the spectrum $\Sigma (A)$ of $A$.
Therefore the r.h.s. of (\rf{pac}) will converge for $z=1$ if there are
no singularities on the segment $[0,1]$, or, if $-1/z\in[-\infty,-1]\not
\in\Sigma (A)$, respectively.
So,
the following Corollary holds.\vst{0.3cm}\nl

\bcl  : Let $A$ be a bounded operator acting in a Banach
space. Let $\Sigma (A)$ be the spectrum of $A$.  If $\Sigma (A)$ is
contained in the complement of $(-\infty ,-1]$, then
$$\frac{1}{(1 + A)} = \int _o^{\infty}
\exp (-\exp t)\  \sum _{n=o}^{\infty}
\frac{(-At)^n}{\mu (n)}\ dt .$$
\lb{mcl3}
\vst{0.3cm}\nl
\ecl
Note that the domain of application of the representation (\rf{pac})
strongly depends on the sign of the parameter $\ld$.
%
%From the physical point of view
%the above restriction on the $\Sigma (A)$ is rather natural
%since in physics one usually works with
%selfadjoint perturbations having bounded spectrum from below.

Now,  to  extend the region of convergence of the
Rayleigh-Schr\"{o}dinger perturbation theory one has to locate the
spectrum of $\lambda VR_o (z)$.
%If  in addition $V$ has relative bound equal
%to zero with  respect to $H_o$ (for instance this is
%the case of the Schr\"{o}dinger  operator  with  the Coulomb
%potential) then due to \ct{44}
The spectrum  of $\lambda VR_o (z)$  for
$z\not\in \Sigma (H_o )$  is  the  set  of  all $\xi = 0$
such that $z\in\Sigma [H(-\lambda/\mu )]$; $\xi = 0$ may or
may not belong to the spectrum \ct{44}. To show this one uses identity
$$
[\ld VR_o(z) -\mu]^{-1} = (-1/\mu)(H_o -z) [H_o+(-\ld/\mu)V -z]^{-1},
$$
for $\mu\neq 0$.
If $z\in\Sigma(H(-\ld/\mu))$ the second factor on the r.h.s. is
not defined. Thus $\mu\in\Sigma(\ld VR_o(z))$. If $z\not\in\Sigma
(H(-\ld/\mu))$ then the second factor is defined and its product
with the unbounded operator $H_o-z$ is a bounded operator on the
whole space.
%, because $V$ has relative bound equal to zero.
%\vst{0.3cm}\nl

{}From  the above discussion and Corollary \rf{mcl3} we have the
Lemma \rf{mlm3}  which converts the problem of covergence of  the
$\mu $-sum for a given $\lambda$ to the problem whether $z$
belongs to the spectrum of some
class of operators $H_{\nu}$ or not.\vst{0.3cm}\nl

\blm : Let $H_o$ and $V$ be as above.
%, and in addition let $V$ have relative bound equal to zero.
Then
$$[1+ \lambda VR_o (z)]^{-1} =\int _o^{\infty}\exp (-\exp t)
          \sum _{n=o}^{\infty} [-\lambda VR_o (z)t]^n/\mu (n) dt$$
          for all z which are not in the spectrum of
      $H_{\nu}$ for $\nu\in\{\lambda u\mid u\in [0,1]\}$.
$\Diamond$\vst{0.3cm}.
\lb{mlm3}
\elm

Note that if $\ld$ is a real number, as usually happens in
physical applications, then one has only to consider
a real parametric family of Hamiltonians in contrast to \ct{44}.
 As a direct consequence of Lemma \rf{mlm3} and (\rf{prt}) one
arrives at the following statement :\vst{0.3cm}\nl
\bcl : If for $\lambda\in\Lambda$ no one of the
branches emanating
from $\varepsilon _o$, since the perturbation parameter
$\nu$ varies on the
straight line  segment  $[0,\lambda ]$,  crosses any other
branches starting from other  eigenvalues in
$\Sigma '(H_o )$, then a
contour $\Gamma$ exists encircling  $\varepsilon _o$  and
$\varepsilon _{\lambda}$ but no other
eigenvalues (see Fig. \rf{fg7}) such that
$$ P_{\lambda} = \int _o^{\infty} \exp (-\exp t)
          \sum _{n=o}^{\infty} R_n
          \frac{(-\lambda t)^n}{\mu (n)}\, dt ,$$
where $R_n$ is the residue of $R_o (z)VR_o (z)...VR_o (z)$ at
$\varepsilon _o$. $\diamondsuit$\vst{0.3cm}
\lb{mth5}
\ecl
%Proof: The only thing to prove is the existence of a contour $\Gamma$
%encircling $\varepsilon_o$ and $\varepsilon_\lambda$ but no other
%eigenvalues without leaving the region of convergence.  But this is
%trivial. $\diamondsuit$

\subsection{Derivative analyticity relations}
\lb{sub:der}

Finally,  we  shall consider the problem of derivative
analyticity  relations (DAR) \cite{BK,KF}. Let us briefly sketch
          the problem.  The  real  and imaginary parts of the forward
          scattering amplitude $F(E)$ in high-energy physics (or in
          optics) are related by a dispersion relation of the form
          \begin{equation}
          Re\ F(E)/E = \frac{2E}{\pi}\ v.p.\ \int_{E^*}^\infty
          \frac{Im\ F(E)}{E(E^2 - E_o^2)} dE,
\lb{r39}
          \end{equation}
          where the pole terms and subtraction constants are for
simplicity omitted. A major ``shortcoming"  of  the  dispersion
          relation is that one has to know the imaginary part $Im\ F(E)$
          on the whole  infinite  integration  interval to obtain the
          real part $Re\ F(E)$  at  a given point. The problem of DAR is
          that of relating the real and imaginary part of $F(E)$ at the
          same point. To cope with  it  in  the general case is quite
    difficult. After the  substitution $x=\ln E$, $x_o =\ln
          E_o$, $k=\ln E^*$ and $f(x)=Im\ F(E)/E$ the integral  on the
 r.h.s. of (\rf{r39}) takes the form,
 $$\frac{1}{\pi} \ v.p.\ \int_{k-x_o}^\infty
\frac{f(x+x_o )}{\sinh (x)}\ dx.$$
          The  solution of the problem of DAR is only known in a
          special  case where $f(x)$ can be analytically extended to an
          entire function \cite{KF,ED}. Using the results of Section
\rf{sec:new}
          one  can establish  the  DAR  for more general class of function
$f(x)$.  In  fact,  under the assumption of regularity of
$f(z)$  on  the  real positive axis one  arrives at  the  following
          representation of the dispersion integral,
\be
\frac{1}{\pi}\ v.p.\  \int_{k-x_o}^\infty dx \int_o^\infty dt
\exp (-\exp t) \sum_{n=o}^\infty  f^{(n)} (x_o )\frac{(xt)^n}{
n!\mu (n)\sinh (x)},
\lb{r40}
\ee
because  of Corollary \rf{rcl1}. If $f(x)$ can be extended to an
          entire function, then $\sum_{n=o}^\infty a_n f^{(n)} (x_o)$
          converges, where
$$a_n = \int_{k-x_o}^\infty dx \int_o^\infty  dt
 \exp (-\exp t)\frac{(xt)^n}{n!\mu (n)\sinh (x)}\cdot$$
One  may  then  integrate term by term in (\rf{r40}), reproducing
the  previous results \cite{KF,ED}. For the references and
recent  status of the problem see also \cite{FK}. A different
question is that of the practical use of the extension because
the real  positive  axis  is  often  the only place where a
scattering amplitude is singular.\np

\section{Conclusion}
\lb{sec:con}

We  have found a family of moment constant summability
methods $\mu _k$ which provide for $k\geq1$ an analytic
continuation of a function regular at the origin onto its
whole  Mittag-Leffler's star, in contrast to the Borel
method (see Remark 2).  These methods are
intimately connected  with the Borel one, since for
$\alpha =\beta =1$ the Borel method is nothing but the
$\mu _o$-method.

The  methods  discussed  above  can be succesfully applied to
{\em convergent}  as well as {\em divergent} perturbation  series
which diverge like $(\ln n)^n$ or slower.
One may encounter such a situation when regularizing a theory on a
lattice \cite{OS,Se} or using a cut-off  in  space time or momentum
space \cite{FMRS1,GJS,S1}.
Of some interest  is
also an expansion  in the ``artificial" parameter proposed
recently by Bender et. al. in the $\lambda\phi ^4_4$ theory
and some other models \cite{67}, and which  seems  to be
convergent.
When some additional information is at disposal or provided
one deals with truncated series our
method can be made more powerful by combineing it with a
conformal mapping or the Pad\'{e} approximation
\cite{26,27,28,CF}. An interesting question  is  to
compare a numerical efficiency  of  the  $\mu _k$-methods
with the  Pad\'{e} or some other methods \cite{BG}.  So far we
did not consider the possible generalization of  the
$\mu _k$-methods to more variables \cite{AT}. Of some interest
is also the question whether
the Wynn  $\varepsilon$-algorithm  for  calculating the Borel
integral could be adapted  to  the $\mu _k$-methods
\cite{Ma}, and whether the inverse of a $\mu _k$-summable
function is $\mu _k$-summable \cite{AM}.

It is obvious that the use of the above methods is not
confined to perturbation theory where the expansion parameter
is a coupling.
The method can be applied to the $1/N$ expansion
\cite{69} or to the $\varepsilon$-expansion
\cite{66} as well.

We  have shown  that the above given summability method
solves the problems ({\bf A}) and ({\bf B}), i.e.,
in the regular case provides an analytic continuation on the
whole Mittag-Leffler (principal) star and
can deal with the horn-shaped  singularity  as well. To our
knowledge it is the {\em first summability method
having these properties}.
We have discussed advantages  and shortcomings of the method,
applications to the Rayleigh-Schr\"{o}dinger perturbation theory
which cover the very important case of the Schr\"{o}dinger
operator in $R^3$ for a system of $s$ particles interacting
with each other by the Coulomb forces \cite{RS,K}, and the
derivative analyticity relations \cite{FK}.

Like  any analytic regular summability method  our
method  may also have a wide domain of applications. We shall not
give a list  of  them  because  the reader can easily judge
whether it is interesting for him or not.

An  open  question  still  remains how the summability
properties (both Borel's and ours) are transported in
equations such as the Dyson-Schwinger equations for Green's
functions \cite{'tH2} and the problem of summing a
perturbation series which violates SAC. To solve the last
problem one has however to have more (nonperturbative) informations
to pick up a physically plausible solution.

One  could  also establish  analogous  results by other
slightly modified Stieltjes moments of the form
          \[\mu (n) := \int _o^\infty\exp (t-\exp t) t^n dt\, ,\]
etc.  In  this  case  one  needs  only  to replace
$\omega (z)$ by
$\tilde{\omega}(z)=\exp z+\ln (z+1)$ and $H_R$ by
$\tilde{H}_R:=\{Re\tilde{\omega}(1/z)>\tilde{\omega}(1/R)\}.$
Note that in contrast to the previous case the boundary
$\partial \tilde{H}_R^{-1}$ of $\tilde{H}_R^{-1}$  approaches
the straight line $Im z=\pi /2$ from below (see Fig. \rf{fg5}).

We hope that a  sufficient number of examples
have been given to illustrate that to draw  physical conclusions
from the Borel (non)summability without knowing the  region of
analyticity to which a power series expansion should be summed
up may sometimes be very dangerous.
We have also shown that to prove the Callan-Symanzik assumption about
the mass
insertion terms one needs generalized SAC, derived in the paper, which
applies to horn-shaped regions.
We then were able to formulate such conditions for a whole variety of
horn-shaped regions.

Finally, our method gives a generalization of \cite{S}
since Theorem \rf{hth2} together with Lemma \rf{hlm5} provide a
summability
mechanism which apart from invariance conditions and linear
covariances  preserves  also  {\em nonlinear perturbative
conditions} such as {\em the unitarity} of the Feynman series.
We  note  that Theorem \rf{hth2} gives a generalization of SAC
in regards to \cite{S,N,So}. Indeed, by the theorem there cannot
exist a function which\vs{0.3cm}\nl
i)  is  analytic  in  the  horn $H_R$ and continuous up to
the boundary;\nl
ii) possesses there the asymptotic expansion (\ref{h10}) which has
equal sign coefficients $a_n$ for $n\geq n_o$.\vs{0.3cm}\nl

We would like to point out that assertions of the type that
a quantity $Q_1$
equals another quantity $Q_2$ to all  orders of
perturbation  theory are very vague unless the SAC are  shown
to  be  valid.
These SAC provide a complete generalization of Simon's  work
\cite{S}. We have given arguments that four dimensional field theories
(without
UV cutoff) violate SAC and that probably there is no reason to look at
the Borel transform of these thories.\vspace{1cm}

\section{Acknowledgements}
First of all I should like to thank my parents for kind attention and
support.
I should also like to acknowledged  friutful discussions
with members of our department.
I am indebted to J. Fischer for continuous interest in
my work  and valuable comments, P. Ho\v{r}ava for discussions on
variety of physical problems, J. Ch\'{y}la,
P. Kol\'{a}\v{r},
S. Neme\v{c}ek, and J. Rame\v{s} for their help with computer
facilities, and Mrs. M. Bou\v{s}kov\'{a} for drawing pictures.

I  take oportunity to thank J. Fuka for discussions about entire
functions which help me in derivation of the properties of
the new summability method in regular case as well as
V. \v{S}ver\'{a}k
for many  stimulating discussions which help me
to derive the analogue of the Nevanlinna theorem for horn-shaped
regions. Discussions with C. Klim\v{c}\'{\i}k on various aspects of
field theories are also greatfully acknowledged.
I thank J.Magnen,
G.'t Hooft, and A.S. Wightman for encouraging correspondence,
CPT, Ecole Polytechnique at Palaisseau for their warm
hospitality and friutful discussions,
E. Seiler for  kind hospitality in M\"{u}nich and discussions,
as well as A. Smith Albion for reading a relevant part of the thesis.
\vspace{1cm}\np

\section{Appendix}
%\abschnitt{Appendix}
\appendix

\noindent Theorems 3.5.1-4 of the Phragmen-Lindel\"{o}f type
\cite{J} can be easily modified to regions  of
the asymptotic type $(1,\eta )$ (and also to other  regions),
          i.e.,
          if $F(z)$

i) is analytic in a region $D$ of the asymptotic type
          $(1,\eta )$,
             and continuous in its closure $\bar{D}$;

          ii)\( \mid F(z)\mid\leq M\exp[-\delta \mid \exp(1/\eta
          z)\mid ] \)\hspace{1in}$z\in \bar{D}$,

\noindent $\delta $ being an arbitrary positive constant, then $F(z)\equiv 0$.

 So,  if  there exist two functions $f_1 (z)$ and
 $f_2 (z)$
 satisfying the hypotheses of the theorem then their difference
 $g(z)=f_1 (z)-f_2 (z)$ obeys the bound
 \[ \mid g(z)\mid\leq A\rho ^n \mid z\mid^n \exp [-\exp w(n)+n\ln w(n)] \]
 uniformly in $z\in \bar{D}$ and $n$. Now, choosing
 %         \begin{equation}
\( n=[w^{-1}(1/\rho r)]=[s(1/\rho r)]\) to optimize the bound,
 %         \end{equation}
 %
where [b] is now entire part of b and $r=\mid z\mid$, one gets that
\[ \mid g(z)\mid\leq A\exp[-\exp(1/\rho r)]
        \hspace{1in}  z\in \bar{D}, \]
for region $D$ of the asymptotic type
$(1,\eta )$ with $\rho\leq\eta $ .
By virtue of ii) then $g(z)\equiv 0$.

On  the other hand, whenever the bound (\rf{eq:b}) is relaxed,
i.e.,  if in a domain $D$ of the  asymptotic  type $(1,\eta )$  a
bound
 \[ \mid R_N (z)\mid\leq A\rho ^N\mid z\mid ^N
      \exp [-\exp w(N)+N\ln w(N)] \hspace{1in} \forall\, z\in\bar{D},\]
is  allowed, with $\rho >\eta $, then for any
$\bar{\rho }$ such that $\rho >\bar{\rho }>\eta$ \
the function $\exp(-\exp(1/\bar{\rho }z))$ satisfies the hypotheses of
the theorem. In fact,
\begin{eqnarray}
 \sup_{z\in\bar{D}}\mid z^{-N}
          \exp(-\exp(1/\bar{\rho} z))\mid =
          \bar{\rho }^N
          \exp\{
         \max_{z\in\overline{1/\bar{\rho }D}}Re[N\ln z-\exp z]\}
          \sim \nonumber\\
\sim\bar{\rho } ^N \exp[-\exp w(N)+N\ln w(N)] .\nonumber
\end{eqnarray}
However,  this  nontrivial  function has trivial asymptotic
series in $D$. $\Diamond$ \np
 \newpage
%{\bf Figure caption}\vspace{0.6cm}
%\listoffigures

\begin{thebibliography}{99}
\bibitem{MS}J. Magnen and R. S\'{e}n\'{e}or, Commun. Math. Phys. {\bf 51}, 297
(1976).
\bibitem{FMRS1}J. Feldman, J. Magnen, V. Rivasseau, and R.
S\'{e}n\'{e}or,
Phys. Rev. Lett. {\bf 54}, 1479 (1985); Commun. Math. Phys. {\bf 103}, 67
(1986); D. Iagolnitzer and J. Magnen, Commun.Math. Phys. {\bf 111}, 81 (1987).
\bibitem{OS}K. Osterwalder and  E. Seiler, Ann. Phys. {\bf 110}, 440 (1978).
\bibitem{1} P. M. Stevenson, Phys.\ Rev.\ D {\bf 23}, 2916 (1981).
\bibitem{2} H. D. Politzer, Nucl.\ Phys.\ {\bf B194}, 493 (1982).
\bibitem{48} I. G. Halliday and P. Suranyi, Phys. Lett.
{\bf 85B} (1979) 421; Phys.\ Rev.\ D {\bf 21}, 1529 (1980).
\bib{Za}V. I. Zakharov, preprint MPI-PAE/PTh 11/91.
\bib{Mo}A. Moroz, Czech. J. Math. {\bf 40}, 200 (1990).
\bibitem{M}A. Moroz, Czech. J. Phys. B {\bf 40}, 705 (1990).
\bibitem{M1}A. Moroz, Commun. Math. Phys. {\bf 133}, 369
(1990).
\bibitem{M2}A. Moroz, Prague preprint PRA-HEP 90/15 (to appear in
Czech. J. Phys. B).
\bibitem{3} F. J. Dyson, Phys. Rev. {\bf 85}, 631 (1952).
\bibitem{4} C. A. Hurst, Proc. Camb. Phil. Soc. {\bf 48}, 625
          (1952); A. Petermann, Helv. Phys. Acta {\bf 26}, 291 (1953);
          W. Thirring, Helv. Phys. Acta {\bf 26}, 33 (1953).
\bibitem{5} L. N. Lipatov, ZhETF {\bf 72}, 411 (1977) (in
          Russian) /Sov. JETP {\bf 45} (1977)/.
\bibitem{BGZ} E. Br\'{e}zin, J. C. Le Guillou, and J.
 Zinn-Justin, Phys.\ Rev.\ D {\bf 15}, 1544 (1977).
\bibitem{P}  G. Parisi, Phys.\ Lett.\ {\bf 66B}, 382 (1977).
\bibitem{9}  C.  M.  Bender and T. T. Wu, Phys.\ Rev.\ {\bf
184}, 1231 (1969).
\bib{GGS}S. Graffi, V.  Grecchi, and  B.  Simon,
Phys.\ Lett.\ {\bf 32B}, 631 (1970).
\bibitem{AM1} G. Auberson and G. Mennessier, Phys.\ Lett.\ {\bf 126B},
263 (1983).
\bibitem{11} B. L. Ioffe, Dokl. Akad. Nauk SSSR {\bf 94}, 437
(1954) (in Russian).
\bibitem{'tH1}G. 't  Hooft, in: {\em The whys of subnuclear
physics}. Proceedings of Erice
Summer School 1977, ed. A. Zichichi (Plenum  Press, New  York, 1979)
p. 943.
\bibitem{GP} D. Gross and V. Periwal, Phys. Rev. Lett.
{\bf 60}, 2105 (1988); Phys. Rev. Lett. {\bf 61}, 1517 (1988).
\bibitem{55} N. N. Khuri, Phys.\ Rev.\ D {\bf 16}, 1754 (1977);
{\em ibid.} {\bf 20}, 881 (1979).
\bibitem{56} B. Simon, {\em Functional integration and quantum
physics}. New York: Academic Press 1979, Chapt. VI.
\bibitem{60} J. Magnen and V. Rivasseau, Commun. Math. Phys.
{\bf 102}, 59 (1985); J. Magnen, F. Nicolo, V. Rivasseau, and
R. S\'{e}n\'{e}or, Commun. Math. Phys. {\bf 108}, 257 (1987).
\bibitem{61}  F.  David,  J.  Feldman,  and V. Rivasseau,
Commun. Math. Phys. {\bf 116}, 215 (1988).
\bibitem{FMRS}J. Feldman, J. Magnen, V. Rivasseau, and R. S\'{e}n\'{e}or,
Commun. Math. Phys. {\bf 109}, 437 (1987).
\bib{Gr}D. Gross and I. Klebanov, Nucl.\ Phys.\ {\bf B344}, 475 (1990).
\bib{MO}P. Mende and H. Ooguri, Nucl.\ Phys.\ {\bf B339} 641 (1990).
\bib{GM}D. Gross and P. Mende, Phys.\ Lett.\ {\bf 197B}, 129 (1987);
Nucl.\ Phys.\ {\bf B303}, 407 (1988).
\bib{GZ}P. Ginsparg and J. Zinn-Justin, preprint SPhT/90-140;
Phys.\ Lett.\ {\bf 255B}, 189 (1991).
\bibitem{C} E. R. Caianiello, Il Nuovo Cim. {\bf 3}, 223 (1956).
\bib{Br}M. Brown, Class. and Quantum Gravity {\bf 2}, 535 (1985).
\bibitem{14} E. Br\'{e}zin and J. C. Le Guillou, and J.
Zinn-Justin, in {\em Phase Transitions and Critical Phenomena},
edited by C. Domb  and M. S. Green (Academic, New York, 1976),
Vol. VI.
\bibitem{15} K. Wilson, Phys.\ Rev.\ B {\bf 4}, 318 4(1971).
\bib{NR}B. Nickel and J. J. Rehr, J.\ Stat.\ Phys.\ {\bf 61}, 1
(1990).
\bib{GJS}J. Glimm, A. Jaffe, and T. Spencer, Ann. Phys. (NY)
{\bf 101}, 610 (1976); ibid. 631 (1976); D. C. Brydges,
J. Stat. Phys. {\bf 42}, 425 (1986); G. Battle and P. Federbush,
Ann. Phys. (NY) {\bf 142}, 95 (1982); Commun. Math. Phys. {\bf 88},
263 (1983); Lett.\ Math.\ Phys.\ {\bf 8}, 55 (1984); Commun. Math.
Phys. {\bf 109}, 417 (1987); G. Battle, Commun. Math. Phys.
{\bf 94}, 133 (1984); Ann.\ Phys.\ (NY) {\bf 201}, 117 (1990);
P. Federbush, Commun.\ Math.\ Phys.\ {\bf 107}, 319 (1986);
{\em ibid.} {\bf 110}, 293 (1987).
\bib{Sei}E. Seiler, {\em Gauge Theories as a problem of Constructive
Field Theory and Statistical Mechanics}. Lecture Notes in
Physics 159 (Springer, New York, 1982).
\bib{GKu}C. Gruber and H. Kunz, Commun.\ Math.\ Phys.\ {\bf 22}, 133
(1971).
\bib{KP}R. Koteck\'{y} and D. Preiss, Commun. Math. Phys.
{\bf 103}, 491 (1986).
\bib{MP}G. Mack and A. Pordt, Commun. Math. Phys. {\bf 97},
267 (1985); Rev. Math. Phys. {\bf 1}, 47 (1989).
\bibitem{Po} A. Pordt, in:
{\em Nonperturbative quantum field theory}. Proceedings, Cargese
1987, 't Hooft, G. et al. (eds.),
pp. 503-511. New York, London: Plenum Press 1988.
\bibitem{T}H.-J. Timme, Hamburg preprint DESY 89-110.
\bib{CH} S. Coleman and J. Hill, Phys.\ Lett.\ {\bf 159B}, 184 (1985);
R. D. Pisarski and S. Rao, Phys.\ Rev.\ D {\bf 32}, 2081 (1985);
J. D. Lykken, J. Sonnenschein, and N. Weiss, preprint
UCLA/89/ITEP/52; preprint UCLA/90/ITEP/17;
Int.\ J.\ Mod.\ Phys.\ {\bf A6}, 1335 (1991).
\bibitem{64} G. 't Hooft, Nucl.\ Phys.\ {\bf B72}, 461 (1974).
\bibitem{65} J. L. Gervais and  B. Sakita, Phys.\ Rev.\ Lett.\
{\bf 30},
716 (1973), B. Sakita and M. A. Virasoro, Phys.\ Rev.\ Lett.\
{\bf 24}, 1146 (1970).
\bibitem{66} K. Wilson and M. Fischer, Phys.\ Rev.\ Lett.\
{\bf 28}, 240 (1972).
\bibitem{67} C. M. Bender et al., Phys.\ Rev.\ Lett.\ {\bf 58},
2615 (1987);
Phys.\ Lett.\ {\bf 205B}, 493 (1988); Phys.\ Rev.\ D {\bf 38}, 1310 (1988);
Phys.\ Rev.\ D {\bf 39}, 3684 (1989).
\bibitem{68}  E.  Witten,  in  `` {\em Recent developments in
Gauge Theories}", edited by  G. 't Hooft , Cargese 1979.
NATO ASI B59. (New York, Plenum Press 1980) pp. 403-420.
\bibitem{69} E. Witten, Nucl.\ Phys.\ {\bf B149}, 285 (1979); G. G.
Chew, G. Rosenzweig, Phys.\ Rep.\ {\bf 41}C, No.5 (1978); A. Kupiainen,
Commun. Math. Phys. {\bf 73}, 273 (1980).
\bibitem{70}D. J. Gross, A. Neveu, Phys.\ Rev.\ D {\bf 10},
3235 (1974).
\bibitem{71} T. T. Wu, Phys.\ Rep.\ {\bf 49}C, 245 (1979).
\bib{Ka}V. A. Kazakov, Phys.\ Lett.\ {\bf 128B}, 316 (1983).
\bibitem{72} G. Paffuti, P. Rossi, preprint IFUP-TH 16/89.
\bibitem{73} H. Flyvbjerg, preprint NBI-HE-88-24.
\bib{Re}N. Read and S. Sachdev, Nucl.\ Phys.\ {\bf B316}, 609 (1989);
Phys.\ Rev.\ Lett.\ {\bf 66}, 1773 (1991); J. B. Marston and I. Affleck,
Phys.\ Rev.\ B {\bf 39}, 11538 (1989).
\bibitem{24}  G. 't Hooft,  Phys.\ Lett.\ {\bf 119B}, 369
(1982); G. 't Hooft, Phys.\ Rep.\ {\bf 104}C, No.  2-4, 129 (1984);
V. Rivasseau, Phys.\ Lett.\ {\bf 137B}, 98 (1984); C. Kopper, Commun.
Math. Phys. {\bf 116}, 57 (1988).
\bibitem{74} E. Br\'{e}zin, Phys.\ Rep.\ {\bf 49}C, 221(1979).
\bib{FMR}J. Fr\"{o}hlich, A. Mardin, and V. Rivasseau, Commun.
Math. Phys. {\bf 86}, 87 (1982).
\bibitem{75} S. G. Rajeev, Phys.\ Lett.\ {\bf 209B}, 53 (1988);
preprint ER-13065-581; Phys.\ Rev.\ D {\bf 42}, 2779 (1990).
\bibitem{R}  J. F. Ritt, Ann.\ Math.\ {\bf 15}, 18 (1916).
\bibitem{H} G. Hardy, {\em Divergent series} (Oxford Univ.
Press, Oxford (1949).
\bib{S} B. Simon, Phys.\ Rev.\ Lett.\ {\bf 28}, 1145 (1972).
\bibitem{BPST} A. A. Belavin, A. M. Polyakov, A. S. Schwartz,
and Yu. S. Tyupkin, Phys.\ Lett.\ {\bf 59B}, 85 (1975).
\bibitem{Pl} A. M. Polyakov, Nucl.\ Phys.\ {\bf B120}, 429 (1977).
\bibitem{BF}  R.  B.  Bogomolny and  V.  A.  Fateyev, Phys.
Lett. {\bf 71B}, 93 (1977).
\bib{N}F. Nevanlinna, PhD thesis of the Alexander University,
Helsingfors 1918.
\bib{So}A. D. Sokal, J.\ Math.\ Phys.\ {\bf 21}, 261 (1980).
\bibitem{18}  W. Y. Crutchfield, Phys.\ Rev.\ D {\bf 19}, 2370 (1979).
\bibitem{19}  V.  I.  Ogievetsky,  Dokl. Akad. Nauk SSSR
{\bf 109}, 919 (1955) (in Russian) /Sov. Physics-Doklady 1
(1955)/.
\bibitem{20} J. Schwinger, Phys.\ Rev.\ {\bf 82}, 664 (1951).
\bibitem{21} V. S. Popov, V. L. Yeletsky, and A. V. Turbiner,
ZhETF {\bf 74}, 445 (1978) (in Russian) /Sov. JETP
{\bf 47} (1978)/.
\bibitem{34}C.  Itzykson and J.  B.  Zuber,  {\em Quantum  field
theory}. McGraw-Hill Inc. 1980 (in Russian, Mir, Moskva 1984).
\bibitem{22}  S.  Graffi and V.  Grecchi,  Commun. Math.
Phys. {\bf 62}, 83 (1978).
\bibitem{23} J. Avron, I. Herbst, and B. Simon, Commun. Math.
Phys. {\bf 79}, 529 (1981).
\bibitem{26} G.  Baker,  Jr.  et  al.,  Phys.  Rev. Lett.
{\bf 36}, 1351 (1976); J. C. Le Guillou and J.  Zinn-Justin,
Phys.\ Rev.\ Lett.\ {\bf 39}, 95 (1977) and J. Physique Lett.
{\bf 46}, L137 (1985).
\bibitem{27} N. N. Khuri, Phys.\ Lett.\ {\bf 82B}, 83 (1979); S. G.
Gorishnyj, S. A. Larin,  and E. V.  Tkachov,  Phys.\  Lett.\
{\bf 101A}, 120 (1984).
\bibitem{28} G. Parisi, Phys.\ Lett.\ {\bf 69B}, 329 (1977).
 \bibitem{Ko} Ch. Kopper, Max-Planck Institute preprint
MPI-PAE/PTh 25/89.
\bibitem{30} S. Templeton, Phys.\ Lett.\ {\bf B103}, 134 (1981);
Phys.\ Rev.\ D {\bf 24}, 3134 (1981).
\bibitem{SG}G. Sansonne and J. Gerretsen, Lectures on the
Theory of Functions of a Complex Variable, Vol. I
(P. Noordhoff, Ltd. Groningen, The Netherlands 1960) p.431.
\bibitem{35}B.  Lautrup, Phys.\ Lett.\ {\bf 69B}, 109 (1977); G.
Parisi, Phys.\  Lett.\ {\bf 76B}, 65 (1978); G. Parisi, Phys. Rep.
{\bf 49}C, 215 (1979).
\bib{Da}F. David, Nucl.\ Phys.\ {\bf B209}, 433 (1982);
{\em ibid.} {\bf B234}, 237 (1984).
\bibitem{36} K. Gawedzki, A. Kupiainen, and B. Tirozzi, Nucl.
Phys. {\bf B257}, 610 (1985).
\bibitem{37}  C.  de  Calan and V. Rivasseau, Commun. Math.
Phys. {\bf 82}, 69 (1981).
\bibitem{BGZ2}E. Br\'{e}zin, G. Parisi, and J. Zinn-Justin, Phys.
Rev.\ D {\bf 16}, 408 (1977);
G. Parisi, Phys. Lett. {\bf 66B}, 167 (1977).
\bibitem{Z}J. Zinn-Justin, {\em Quantum field theories and critical phenemena}.
(Clarendon Press, Oxford, 1989) ch. 40.
\bibitem{DS}M. Dine and N. Seiberg, Phys.\ Rev.\ Lett.\ {\bf 55}, 366 (1985);
Phys.\ Lett.\ {\bf 162B}, 299 (1985).
\bibitem{BGZ1}E. Br\'{e}zin, J. C. Le Guillou, and J. Zinn-Justin,
Phys.\ Rev.\ D {\bf 15}, 1558 (1977).
\bibitem{32}S. Graffi and V. Grecchi, Phys. Lett. {\bf 121B}, 410
(1983).
\bib{CO}S. Chadha and P. Olesen, Phys.\
Lett.\ {\bf 72B}, 87 (1977); P. Olesen, Phys.\ Lett.\ {\bf 73B}, 327 (1978).
\bibitem{W}D. V. Widder, The Laplace transform
(Princeton Univ. Press, Princeton 1946) Chapt. II.
\bibitem{K1}N. N. Khuri, Phys.\ Rev.\ D {\bf 12}, 2298 (1975).
\bibitem{Wi}A. S. Wightman, in: {\em The Whys of Subnuclear Physics}.
Proceedings of Erice Summer School 1977, ed. A. Zichichi (Plenum  Press,
New  York, 1979) p.983;  Canadian Math. Soc. Conf. Proc. {\bf 9}, 1 (1988).
\bibitem{K0}N. N. Khuri, Phys.\ Rev.\ D {\bf 23}, 2285 (1981).
\bibitem{KR} N. N. Khuri and H. C. Ren, Ann. Phys. {\bf
          189}, 142 (1989).
\bibitem{GK} K. Gawedzki and A. Kupiainen, Nucl. Phys.
{\bf B257}, 474 (1985);
K.Symanzik, Lett.\ Nuovo Cim.\ {\bf 6}, 77 (1973).
\bibitem{Sy}K. Symanzik, Commun. Math. Phys. {\bf 18}, 227 (1970);
{\em ibid.} {\bf 23}, 49 (1971).
\bibitem{59} M. C. Berg\'{e}re and F. David, Phys.\ Lett.\
{\bf 135B}, 412 (1984).
\bibitem{63}  M. Aizenman, Commun.\ Math.\ Phys.\ {\bf 86}, 1 (1982);
J. Fr\"{o}hlich, Nucl.\ Phys.\ {\bf B200}, 281 (1984).
\bib{LW}M. L\"{u}scher and P. Weisz, Nucl.\ Phys.\ {\bf B290} [FS 20],
25 (1987); M. L\"{u}scher, Nucl. Phys. {\bf B300} [FS 22], 325 (1990);
N. V. Krasnikov, Mod. Phys. Lett. {\bf A6}, 693 (1991).
\bibitem{57} C. G. Callan, R. Dashen, and D. J. Gross, Phys.\
Lett.\ {\bf 63B}, 334 (1976); W. E. Caswell, Ann.\ Phys.\ {\bf 123},
153 (1979).
\bibitem{58} S. Graffi and V. Grecchi, Phys. Lett. {\bf 121B},
410 (1983).
\bib{F}M. V. Fedoriuk, {\em Asymptotic estimates: Integrals and series}
(in Russian) (Nauka, Moscow, 1987).
\bibitem{J}M. A. Evgrafov,  {\em Asymptotic  estimates and entire
functions} (in Russian) (Nauka, Moscow, 1979).
\bib{E}A. Erd\'{e}lyi (ed.), {\em Higher
transcendental functions}, Vol. III.
(McGraw-Hill, New York, 1955) Chapter XVIII.
\bib{L}E. Lindel\"{o}f, Bull.\ Sci.\ Math.\ {\bf 27}, 213 (1903).
%\np
\bibitem{RS} M. Reed and B. Simon, {\em Methods of modern
mathematical physics}, Vol. IV., {\em Analysis of operators}
(Academic Press, New York, 1978).
\bibitem{K}T. Kato, {\em Perturbation theory for linear
operators}, 2-nd edition (Springer, New York, 1976).
\bib{GGM}E. Calicetti, V. Grecchi, and M. Maioli, Commun.
Math. Phys. {\bf 104}, 163 (1986).
\bibitem{44}M. Reeken, J. Math. Phys. {\bf 11}, 822 (1970).
\bib{BK}J.  B.  Bronzan,  G. L. Kane, and U. P. Sukhatme,
Phys.\ Lett.\ {\bf 49B}, 272 (1974).
\bib{KF}P. Kol\'{a}\v{r} and J. Fischer, J.\ Math.\ Phys.\ {\bf 25}, 2538
(1984).
\bib{ED}G. K. Eichmann and J. Dronkers, Phys.\ Lett.\ {\bf 52B}, 428 (1974).
\bib{FK}J. Fischer and P. Kol\'{a}\v{r}, Czech. J.\ Phys.\ B {\bf 37}, 297
(1987).
\bib{S1}B. Simon, Il Nuovo Cim. {\bf 59}A, 199 (1969).
\bib{Se}N. Seiberg, Phys. Rev. Lett. {\bf 53}, 637 (1984).
\bib{CF}S. Ciulli and J. Fischer, Nucl.\ Phys.\ {\bf 24}, 465 (1961).
\bib{BG}G. Baker, Jr., and P. Graves-Morris, {\em
Pad\'{e} approximants}, Vol. I. (Adison-Wesley, London, 1981).
\bib{AT}  L.  A.  Aizenberg and V. N. Trutnev,
Sib. Math. J. {\bf 12}, 1398 (1971),
G. Parisi, Phys.\ Rep.\ {\bf 49}C,  215 (1979).
\bib{Ma}I. O. Mayer, Theor. Math. Phys. {\bf 75}, 234 (1988) (in Russian)
 /Sov. JETP 75 (1988)/.
\bib{AM}G. Auberson and G. Mennessier, Commun. Math. Phys.
{\bf 100}, 439 (1985),
W. Boenkost, C. Bervillier, and V. Scharffenberger, J. Math. Phys. {\bf 29},
1118 (1988).
\bibitem{'tH2}G. 't Hooft, G.: Private correspondence.
%\bibitem{53} W. Y. Crutchfield, Phys.\ Lett.\ {\bf 77B}, 109 (1978).
%\bibitem{AM}G. Auberson and G. Mennessier, J. Math. Phys. {\bf 22} (1981)
%2472.
%\bibitem{M3}A. Moroz, in preparation.
\end{thebibliography}
\end{document}